\documentclass[showpacs,pre,aps,twocolumn,superscriptaddress,longbibliography]{revtex4-1}

\usepackage{amsmath,amsfonts,amssymb,bm,graphicx,hyperref,mathrsfs}
\usepackage{color}

\usepackage{natbib}

\usepackage{float}

\usepackage{ulem}

\usepackage[subpreambles=true]{standalone}

\usepackage[percent]{overpic}
\usepackage{etoolbox}

\newcommand{\ee}{\mathrm{e}}

\newcommand{\beq}{\begin{equation}}
\newcommand{\eeq}{\end{equation}}

\newcommand{\bP}{\bm{P}}
\newcommand{\bJ}{\bm{J}}
\newcommand{\bF}{\bm{f}}

\newcommand{\rtptime}{\tau_{\rm p}}

\newcommand{\nubar}{\overline{\nu}}
\newcommand{\bnu}{\bm{\nu}}

\newcommand{\lp}{\tilde{l}_{\rm p}}

\definecolor{mygreen}{rgb}{0.0,0.4,0.25}

\newcommand{\xp}{51}
\newcommand{\yp}{48}

\begin{document}

\title{Collective motion in large deviations of active particles}

\author{Yann-Edwin Keta}
\affiliation{Department of Applied Mathematics and Theoretical Physics, University of Cambridge,
Wilberforce Road, Cambridge CB3 0WA, United Kingdom}
\affiliation{Universit\'e Paris Diderot, Laboratoire Mati\`ere et Syst\`emes Complexes (MSC),
UMR 7057 CNRS, F-75205 Paris, France}
\affiliation{D\'epartement de Physique, \'Ecole normale sup\'erieure de Lyon,
69364 Lyon Cedex 07, France}
\author{{\'E}tienne Fodor}
\affiliation{Department of Applied Mathematics and Theoretical Physics, University of Cambridge,
Wilberforce Road, Cambridge CB3 0WA, United Kingdom}
\author{Fr{\'e}d{\'e}ric van Wijland}
\affiliation{Universit\'e Paris Diderot, Laboratoire Mati\`ere et Syst\`emes Complexes (MSC),
UMR 7057 CNRS, F-75205 Paris, France}
\author{Michael E. Cates}
\affiliation{Department of Applied Mathematics and Theoretical Physics, University of Cambridge,
Wilberforce Road, Cambridge CB3 0WA, United Kingdom}
\author{Robert L. Jack}
\affiliation{Department of Applied Mathematics and Theoretical Physics, University of Cambridge,
Wilberforce Road, Cambridge CB3 0WA, United Kingdom}
\affiliation{Department of Chemistry, University of Cambridge, Lensfield Road, Cambridge CB2 1EW, United Kingdom}

\begin{abstract}
We analyse collective motion that occurs during rare (large deviation) events in systems of active particles, both numerically and analytically.
We discuss the associated dynamical phase transition {to collective motion}, which occurs when the active work is biased towards larger values, {and is associated with alignment of particles' orientations}.
A finite biasing field is needed to induce spontaneous symmetry breaking, even in large systems.
Particle alignment is computed exactly for a system of two particles.  For many-particle systems, we analyse the symmetry breaking by
an optimal-control representation of the biased dynamics,
and we propose a fluctuating hydrodynamic theory that captures the emergence of polar order in the biased state.
\end{abstract}

\maketitle

\section{Introduction}

\subsection{Motivation}

Active matter emerged in the last decades as a novel class of nonequilibrium soft systems where every constituent consumes and dissipates energy to produce a self-propelled motion~\cite{Marchetti2013, Bechinger2016, Marchetti2018}. It includes both living and social systems, such as swarms of bacteria~\cite{Libchaber2000, Elgeti2015} bird flocks~\cite{Cavagna2010, Cavagna2014} and human crowds~\cite{Bottinelli2016, Bartolo2019}, as well as synthetic systems, such as vibrated particles~\cite{Dauchot2010, Sood2014} and self-catalytic colloids in a fuel bath~\cite{Palacci2013, Bechinger2013}. In these experimental systems, the combination of self-propulsion and interaction can lead to collective behavior without any equilibrium equivalent. Collective motion with orientational order~\cite{Dauchot2010, Sood2014} and the spontaneous formation of particle clusters despite the absence of attractive interactions~\cite{Palacci2013, Bechinger2013} are celebrated examples.

Minimal models have been proposed to capture these collective effects, with a view to identifying the essential ingredients of the dynamics which delineate generic classes of active matter. The emergence of collective motion (CM) is generally described by the Vicsek model in terms of aligning active particles~\cite{Vicsek1995}, and its equivalent Toner-Tu model at hydrodynamic level~\cite{Toner1995, Chate2020}, whereas purely repulsive active particles yielding a motility-induced phase separation (MIPS) are usually considered to reproduce the behavior of isotropic self-propelled particles~\cite{Fily2012, Redner2013, Cates2015}. To characterize the structure and dynamics, thermodynamic tools inspired by equilibrium have been proposed, such as pressure~\cite{Marchetti2014, Brady2014, Solon2015b}, others focus specifically on the deviation from equilibrium, such as the irreversibility of the dynamics~\cite{Nardini2016, Mandal2017, Pietzonka2017, Nardini2017, Shankar2018, Bo2019} and the dissipation of energy~\cite{Toyabe2010, Speck2016, Ahmed2016, Tociu2019, Nemoto2020}.

Several recent studies focused on large deviations of active matter~\cite{Suma2017, Grandpre2018, Whitelam2018, Nemoto2019, Tociu2019, Gradenigo2019, Mallmin2019, Nemoto2020, Cagnetta2020, Suma2020, Grandpre2020}. They consider transient rare events where the system does not behave ergodically. Such events are often  accompanied by collective effects, and may also lead to dynamical phase transitions, where atypical trajectories differ significantly from the typical ones~\cite{Lecomte2007, Touchette2009, Jack2010, Jack2015, Jack2020}. Numerical techniques can be used to analyse these transient events by introducing a bias parameter which controls the distance from the typical dynamics~\cite{Kurchan2006, Nemoto2016}. They open the door to studying the microscopic mechanisms leading to stabilize atypical collective behaviors. These techniques have already proved successful to unveil dynamical transitions in glassy dynamics~\cite{Garrahan2007, Hedges2009, Speck2012} and high-dimensional chaotic chains~\cite{Tailleur2007, Laffargue2013}.

It was recently shown~\cite{Nemoto2019} that some large deviations of isotropic active particles are associated with CM.  In this  dynamical transition, long-ranged orientational order is stabilized, despite the absence of any microscopic interactions that favour alignment.
This stands in contrast to the usual (and intuitive) expectation that CM emerges as a result of particle alignment.
However, that work did not resolve the nature of the transition between the isotropic and CM phases.  This work develops further our understanding of this transition, including the mechanism of spontaneous symmetry breaking, the location of the phase transition, and the relationship of the CM with the hydrodynamic dynamics of the system.

\subsection{Summary of main results}

Before describing our analysis, we summarise the main results. We consider {active Brownian particles (ABPs) as a popular model of overdamped self-propelled particles~\cite{Fily2012, Redner2013}. For a long time interval of duration $\tau$, we focus on the time-averaged rate of the active work per particle $w_\tau$, which quantifies how much the self-propulsion forces of particles translate into actual displacement.}
The ensemble-averaged rate is $\langle w_\tau \rangle$.
Full definitions are given in Sec.~\ref{sec:methods}, below.

\begin{figure}
\centering
\includegraphics[width=0.48\textwidth]{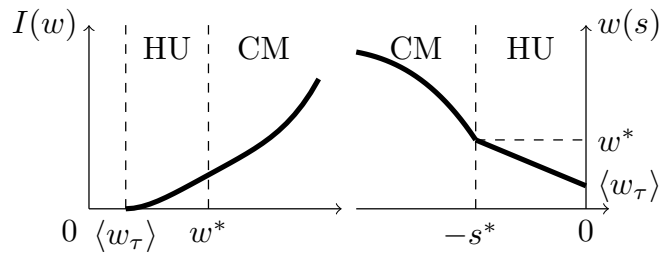}
\caption{Schematic behaviour of {\bf (left)} the rate function, following  Fig.~1(a) of~\cite{Nemoto2019} and Fig.~\ref{fig:rate}, and {\bf (right)} the active work $w_{\tau}$ as a function of its conjugate field $s$, following Fig.~1(b) of~\cite{Nemoto2019} and Figs.~\ref{fig:ws-nu},~\ref{fig:fss}.
Vertical dashed lined delimit the two regimes identified: {CM} $\equiv$ collectively moving state, {HU} $\equiv$ isotropic hyperuniform state.  For positive $s$ (or equivalently $w<\langle w_\tau\rangle$), the system is phase-separated and arrested~\cite{Nemoto2019}, that case is not discussed here.
}%
\label{fig:summary}
\end{figure}

Building on~\cite{Nemoto2019}, we focus on large deviations where the active work is enhanced.
The resulting picture is summarised in Fig.~\ref{fig:summary}, as a function of the active work $w$, and also its conjugate field $s$.
We restrict to situations where the steady state of the system is spatially homogeneous, so the activity of the particles is not enough to cause  MIPS.
We find that spontaneous symmetry breaking occurs for values of the active work $w_\tau$ beyond a threshold $w^*$ that is strictly greater than its average value $\langle w_\tau \rangle$.  There is a corresponding threshold for the biasing field, in that collective motion takes place for $s<-s^*$ (this sign convention is chosen so that $s^*>0$, it means that the transition takes place at $s=-s^*$ and not $s=s^*$).
This result is supported by a finite-size scaling analysis.
It resolves an open question from~\cite{Nemoto2019}, as to whether symmetry breaking might be present for all $w_\tau > \langle w_\tau \rangle$, in sufficiently large systems.
For ABPs, an important parameter is the rotational diffusion constant $D_r$ which determines the correlation time of the self-propulsion force.
We find that $s^* \sim D_r$ for small $D_r$, which is the regime where the system differs strongly from a passive fluid.
A consequence of this analysis is that the system is an isotropic fluid phase for $-s^* < s < 0$.  Generic arguments~\cite{Jack2015,Dolezal2019,Jack2020} based on coupling between large deviations and hydrodynamic modes mean that this phase is hyperuniform~\cite{Torquato2003}.
(The CM phase may also be expected to have a similar property but that question is not addressed here.)

To analyse the mechanism of symmetry breaking, we first solve exactly a system of two active run-and-tumble particles (RTPs), to demonstrate that particles naturally align during large deviation events.
Turning to many-particle systems, we exploit connections between large deviation theory and optimal control theory~\cite{Touchette2015, Jack2015, Jack2020}, and we also develop a Landau-Ginzburg theory for the symmetry-breaking transition, which includes both the orientational order of the ABPs, and their hydrodynamic density fluctuations.  This gives a detailed description of the CM phase.

The structure of the of the paper is as follows: Sec.\ref{sec:methods} describes the models and outlines the theoretical background;
Sec.~\ref{sec:symm-breaking} presents numerical results for CM; Sec.~\ref{sec:two-rtps} analyses the two-particle case;
Sec.~\ref{sec:mechanism} explores the mechanism using
 large-deviation bounds based on controlled systems with orientational interactions;
 Sec.~\ref{sec:landau} describes a {Landau-Ginzburg} theory for the CM transition.
 Conclusions are summarised in Sec.~\ref{sec:conclusion} and several appendices contain additional technical information.

\section{Model and methods}
\label{sec:methods}

\subsection{Active Brownian particles}

We consider $N$ active Brownian particles (ABPs) in two spatial dimensions \cite{Nemoto2019}. Their positions and orientations are $\bm{r}_i$ and $\theta_i$. We define an orientation vector $\bm{u}(\theta_i) = (\cos\theta_i, \sin\theta_i)$ which we sometimes abbreviate simply as $\bm{u}_i$.  The particles are self-propelled with (bare) speed $v_0$, they interact through a WCA interaction potential $V(r)$ with range $\sigma$ and strength $\varepsilon_0$.   Define
\beq
U = \frac{1}{T} \sum_{1\leq i<j\leq N} V(|\bm{r}_i-\bm{r}_j|)
\eeq
as the dimensionless potential energy, which has been rescaled by the temperature $T$.
We take Boltzmann's constant $k_{\rm B} = 1$. 
The equations of motion are
\beq
\begin{aligned}
\dot{\bm{r}}_i & = v_0 \bm{u}(\theta_i) -
D \nabla_i U
+ \sqrt{2D} \, \bm{\eta}_i \;\\
\dot{\theta}_i & = \sqrt{2 D_r} \, \xi_i,
\end{aligned}
\label{equ:dr}
\eeq
where $\bm{\eta}_i$, $\xi_i$ are zero-mean unit-variance Gaussian white noises,
and $D$, $D_r$ are translational and rotational diffusivities.
The combination $D\nabla_i U$ in (\ref{equ:dr}) should be interpreted as the product of a mobility $\mu$ and the gradient of potential energy; here we have used $\mu = D/T$ by the fluctuation-dissipation theorem.

For consistency with Refs.~\cite{Redner2013,Nemoto2019}, we set $D_r = 3D/\sigma^2$ in accordance with the Stokes-Einstein-Debye relation.
The particles are contained in a periodic box of size $L\times L$, the dimensionless measure of density is $\phi = N \pi \sigma^2/(4 L^2)$.
For numerical work we consider a single density $\phi =0.65$, results for other densities are similar~\cite{Nemoto2019}.
We also take $ T=\varepsilon_0$.

For a single isolated particle, the effect of the self-propulsion force is that the particle follows a persistent random walk with persistence length $l_{\rm p} = v_0/D_r$.  Dividing this length by the particle diameter defines an important
dimensionless control parameter
\beq
\lp = \frac{ v_0 }{ \sigma D_r }
\eeq
which determines the effect of the active self-propulsion.   For $\lp\to0$ we have a passive (equilibrium) system.  For large $\lp\gtrsim 15$ the self-propulsion leads to motility induced phase separation (MIPS)~\cite{Cates2015}.
We note that since $D\propto D_r$, increasing $\lp$ changes the balance between the self-propulsion term and the repulsive (WCA) forces in (\ref{equ:dr}).  This means that large $\lp$ tends to make particles overlap more -- they appear to be softer.

When presenting numerical results, we take $\sigma=1$ as the unit of length and we fix the time unit by setting also $v_0=1$.  For theoretical calculations{, we} retain $v_0$ and $\sigma$ as explicit quantities.

\subsection{Dissipation and active work}

We define the instantaneous dissipated power from a purely mechanical argument as the rate of work that the particles exert on the solvent~\cite{Sekimoto1998, Seifert2012}
\beq
\dot{\mathcal{W}} = \sum_{i} \dot{\bm{r}}_i \circ \frac{1}{D} \left(\dot{\bm{r}}_i - \sqrt{2D} \bm{\eta}_i\right)
\eeq
where $\circ$ is a Stratonovich product.  We have absorbed a factor of $T$ into ${\cal W}$, to obtain a reduced (dimensionless) work.  Here and in the following, sums run over all particles, unless otherwise stated.
Using (\ref{equ:dr}) and taking a time average, we find
\beq
\begin{aligned}
\frac{1}{\tau} \int_0^{\tau} \dot{\mathcal{W}}(t) \, \mathrm{d}t
=  \frac{N v_0^2}{D}  w_{\tau}  + \frac{1}{\tau} [ U(\tau) - U(0) ]
\label{equ:work-heat}
\end{aligned}
\eeq
where
\beq
w_\tau = \frac{1}{v_0 N\tau} \sum_{i} \int_0^\tau \bm{u}(\theta_i) \circ \mathrm{d}\bm{r}_i
\label{equ:def-w}
\eeq
is the (reduced) active work per particle~\cite{Nemoto2019}.  This is a natural measure of how efficiently active forces create motion.
It is normalised such that $\langle w_{\tau} \rangle = 1$ in the dilute limit $\phi \to 0$, while $\langle w_{\tau} \rangle = 0$ for a completely jammed system.  For a steady state, the term involving $U$ in (\ref{equ:work-heat}) is zero on average, so the average dissipation is fully determined by the average of the active work.

This active work $w_{\tau}$ is also related to the entropy production rate in the full $\{\bm{r}_i, \theta_i\}$ configuration space (which considers self-propulsion as a quantity that is even under time-reversal\cite{Nemoto2019, Shankar2018}).   This differs in general from the entropy production measured in position space $\{\bm{r}_i\}$ \cite{Nardini2016, Puglisi2017, Marconi2017}.

Since (\ref{equ:dr}) has three separate contributions, there is a natural decomposition of the active work
\beq
\label{equ:wdecomp}
w_{\tau} = 1 + w_{f,\tau} + w_{\eta,\tau}
\eeq
where the constant term stems from the product of the self-propulsion direction with itself and
\begin{align}
\label{equ:wf}
w_{f,\tau} &= \frac{-D}{v_0 N \tau} \sum_{i} \int_0^{\tau} \bm{u}(\theta_i) \cdot \nabla_i U  \, \mathrm{d}t,\\
\label{equ:weta}
w_{\eta,\tau} &= \frac{1}{v_0 N \tau} \sum_{i} \int_0^{\tau} \bm{u}(\theta_i) \circ \sqrt{2 D} \bm{\eta}_i\, \mathrm{d}t \; .
\end{align}
On average $\langle w_{\eta,\tau} \rangle = 0$, so $\langle w_{\tau} \rangle = 1 + \langle w_{f,\tau} \rangle$.
The quantity $w_{f,\tau}$ is negative on average, because collisions between particles tend to involve particle orientation vectors being anti-parallel to the interparticle force.  In the following, we consider situations where the system self-organises to reduce collisions, in which case $w_{f,\tau}$ becomes less negative (it increases towards zero).

\subsection{Large deviations}
\label{sec:large-deviations}

For any given $N$, the active work $w_{\tau}$ satisfies a large deviation principle in the limit of large $\tau$~\cite{Nemoto2019}
\beq
p(w_{\tau}) \asymp \exp\left[- \tau N I(w_{\tau})\right]
\eeq
where $I(w_{\tau})$ is a scaled rate function. We define the scaled cumulant generating function (SCGF)
\beq
\label{equ:scgf}
\psi(s) = \lim_{\tau \to \infty} \frac{1}{N \tau} \log \left\langle \exp\left(- s N \tau w_{\tau} \right) \right\rangle,
\eeq
related to $I(w_{\tau})$ by Legendre transformation (see Eq.~\ref{equ:I-legendre} below), and where we have introduced  a biasing parameter $s$.
The SCGF can be obtained by solving an eigenvalue problem, see Appendix~\ref{app:abp-LD}.

There is a useful analogy between this dynamical large deviation formalism and equilibrium statistical mechanics.  We recall the central features of this analogy, see also~\cite{Touchette2009, Lecomte2007,Jack2020}. Trajectories of our $2$-dimensional system are analogous to configurations of a $2+1$-dimensional system.
Also, the biasing field $s$ corresponds to a thermodynamic field conjugate to the active work.  The SCGF $\psi(s)$ corresponds to the free energy density, and is thus sometimes referred to as the dynamical free energy. Any singularity in this function is {a signature of} a dynamical phase transition.  In particular, we focus below on phase transitions where rotational symmetry of the system is spontaneously broken.

Continuing with this analogy, the average in (\ref{equ:scgf}) corresponds to a partition function for Boltzmann-like distribution of trajectories.
Averages with respect to this distribution take the form
\beq
\left\langle\mathcal{A}\right\rangle_s = \frac{\left\langle\mathcal{A} \, e^{-s N \tau w_{\tau}}\right\rangle}{\left\langle e^{-s N \tau w_{\tau}} \right\rangle}
\label{equ:ave-bias}
\eeq
where ${\cal A}$ is a dynamical observable.
Numerical computation of such averages is challenging in general -- it is comparable to computing a thermodynamic average at temperature $T$ by reweighting from an equilibrium system with temperature $T'\neq T$.  To achieve this,
we evolve simultaneously a large population of copies of the system to generate ``biased ensembles'' by cloning and deleting some of these copies at regular steps in order to enforce the dynamical effective Boltzmann distribution. This method, known as a cloning algorithm \cite{Kurchan2006,Lecomte2007-cloning}, allows estimation of averages like
(\ref{equ:scgf},\ref{equ:ave-bias}) with a cost that scales linearly in $\tau$, allowing direct access to the large-$\tau$ limit.
We implement it following~\cite{Nemoto2016,Brewer2018}, using a modified equation of motion to evolve the clones. Details are given in  Appendix~\ref{app:cloning}.

\begin{figure}
\centering
\includegraphics[width=0.48\textwidth]{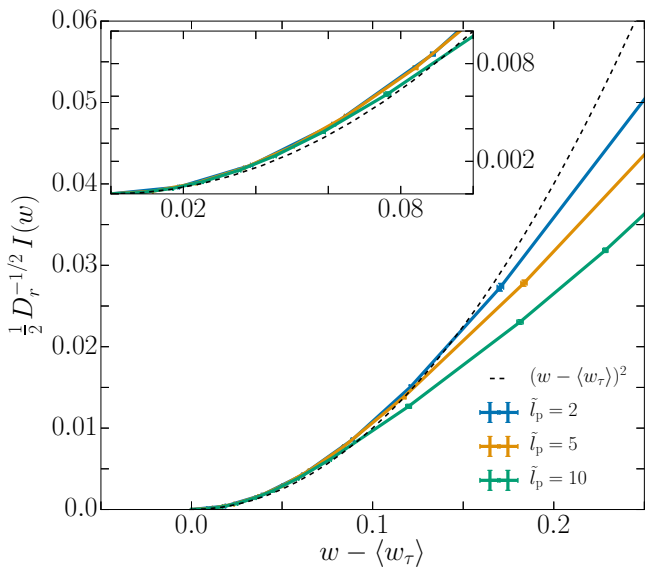}
\caption{Rate function $I(w)$ computed with (\ref{equ:I-legendre}) rescaled by $D_r^{1/2}$ for persistence lengths $\lp = 2, 5, 10$.
The inset is a magnified version of the behavior for small $w-\langle w_\tau\rangle$. \textit{Parameter values}: $N=50$, $\phi=0.65$, $n_c = 10^3$, $t_{\mathrm{max}} = 10^3$.}
\label{fig:rate}
\end{figure}

Of particular interest is the quantity
\beq
w(s) = \lim_{\tau\to\infty} \langle w_\tau \rangle_s
\eeq
which obeys $w(s) = -\psi'(s)$.  Since $\psi$ is convex \cite{Touchette2009}, this is a decreasing function of $s$, we denote its inverse by $s(w)$.
The rate function $I$ is related to the SCGF by Legendre transform, in particular
\beq
I(w) 
= - w s(w)  - \psi(s(w))
\label{equ:I-legendre}
\eeq
which allows computation of the rate function from the output of a cloning simulation.

Fig.~\ref{fig:rate} shows the rate function of the active work $w\geq \langle w_\tau\rangle$, for different persistence lengths $\tilde{l}_p$.
The vertical axis has been scaled by $D_r^{-1/2}$ which leads to data collapse near the minimum.
The rate function is minimal (and equal to zero) at $w=\langle w_\tau\rangle$ and its curvature there is related to the variance of the active work as
\beq
\frac{1}{ I''( \langle w_\tau\rangle ) } =  \lim_{\tau \to\infty} \tau N [ \langle w_\tau^2 \rangle - \langle w_\tau \rangle^2 ]
\label{equ:I''}
\eeq
Our data suggest that this variance is proportional to $D_r^{-1/2}$.  As $w$ increases from $\langle w_\tau \rangle$, the rate function deviates from a quadratic form, in particular its curvature decreases, showing that large fluctuations of $w_\tau$ are less unlikely than a simple Gaussian approximation would predict.  This is related to a dynamical phase transition, as we now explain.

\section{Evidence for symmetry breaking}
\label{sec:symm-breaking}

\subsection{Collective motion and symmetry breaking}
\label{sec:symm-break-intro}

Ref.~\cite{Nemoto2019} focussed on a system whose parameters lie (as $N \to \infty$) within the MIPS region, $\lp = 40$ and $\phi = 0.65$ (see Ref.~\cite{Redner2013} for the full phase diagram of the system). For $s > 0$, {\it i.e.} biasing towards trajectories of low active work, those trajectories involve a coexistence of a dense jammed, arrested domain with a dilute vapor, given the name phase-separated arrest (PSA). For $s < 0$, {\it i.e.} biasing towards trajectories of high active work, collective motion (CM) is found with aligned propulsion directions, despite the absence of aligning interactions microscopically.
In this work, we consider exclusively trajectories with positive fluctuations of the active work ($s < 0$). Compared to~\cite{Nemoto2019}, we focus on lower persistence lengths $\lp$, such that the unbiased behaviour of the system is that of an homogeneous active fluid.

The physical reason for CM when {$s<0$} is that if particles all travel in the same direction with speed $v_0$, they collide much less frequently, so $w_{f,\tau}$ is increased.  In particular, if the $\bm{u}_i$ are random unit vectors then there are large relative velocities between particles (because $|\bm{u}_i-\bm{u}_j|$ is typically of order unity).  On the other hand, perfectly aligned orientations lead to $|\bm{u}_i-\bm{u}_j|=0$, so the only sources of relative motion are the passive noises $\bm{\eta}_i,\bm{\eta}_j$.  The larger the relative velocities, the more often the particles collide, leading to smaller (more negative) values of $w_{f,\tau}$.  Hence collective motion is a natural mechanism for increasing $w_\tau$.

It is also notable that $w_\tau$ is closely related to the ratio $v(\rho)/v_0$ that appears in theories of MIPS~\cite{Solon2015b}, and measures the reduction in particle speed due to collisions.  This further emphasises that larger active work corresponds to reduced collisions.

Since the CM phase is associated with spontaneous breaking of rotational symmetry, it is natural to identify an
order parameter,
\beq
\bm{\nu} = \frac{1}{N} \sum_{i} \bm{u}_i \; .
\label{equ:def-nu}
\eeq
The particle orientations $\bm{u}_i$ evolve independently of their positions, so the steady state distribution of $\bm{\nu}$ is simply the distribution of the average of $N$ random unit vectors.
It is convenient to define also the time-average of the modulus of the order parameter:
\beq
\nubar_\tau = \frac{1}{\tau} \int_0^\tau |\bm{\nu}(t)| dt \; .
\label{equ:bar-nu-tau}
\eeq
For large times one has $\langle \nubar_\tau \rangle_s = \langle |\bm{\nu}| \rangle_s$.

For large $N$, the central limit theorem means that $p(\bm{\nu})\to (N/\pi)\ee^{-N|\bm{\nu}|^2}$ (in distribution) and hence
\beq
\langle \nubar_\tau \rangle
\simeq \frac12 \sqrt{\frac{\pi}{N}}
\label{equ:para}
\eeq
which tends to zero as $N\to\infty$.
On the other hand, symmetry-broken states have
\beq
\langle \nubar_\tau \rangle_s
= O(1)
\label{equ:ferro}
\eeq
as $N\to\infty$.

As $N\to\infty$, the limiting value of $\langle \nubar_\tau \rangle_s$ is zero throughout the isotropic phase,
but non-zero in the CM phase.  This leads to a singularity at the transition point $s=s^*$, as expected for an order parameter.
However, in finite systems, the quantity $\langle  | \bm{\nu} | \rangle_s$  is always positive and has a smooth (analytic)
dependence on the field $s$.
To identify the phase transition in numerical studies, we use that the
the finite-size scaling behaviour (\ref{equ:para},\ref{equ:ferro}) is different in the two phases.

\subsection{Results for $\lp\geq2$}
\label{sec:results-symm}

\begin{figure}
\centering
\begin{overpic}[trim=0 0 0 0, width=0.48\textwidth]{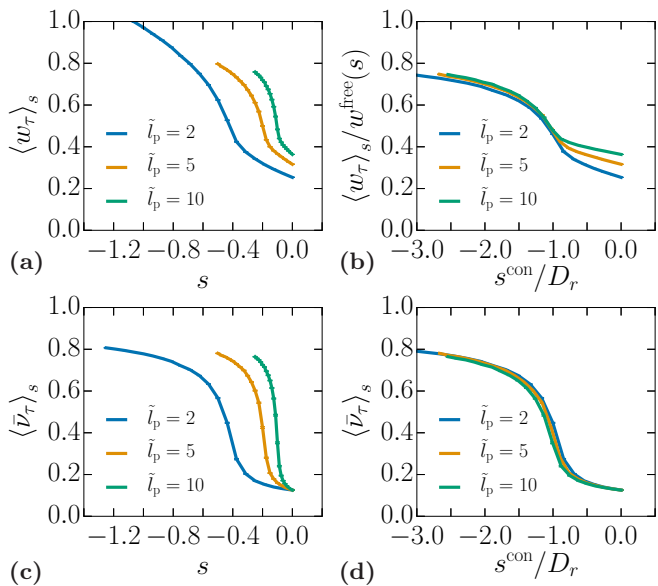}
\put (0, \yp) {\texttt{\bf (a)}}
\put (\xp, \yp) {\texttt{\bf (b)}}
\put (0, 0) {\texttt{\bf (c)}}
\put (\xp, 0) {\texttt{\bf (d)}}
\end{overpic}
\caption{{\bf (a)} Biased average of the active work $\langle w_{\tau} \rangle_s$.
{\bf(b)}~The data from panel (a) plotted using rescaled variables from (\ref{equ:w-free},\ref{equ:s_con}), leading to data collapse.
 {\bf (c)}~Biased average of the polarisation norm $\langle \bar{\nu}_{\tau} \rangle_s$ (See Eq.~\ref{equ:bar-nu-tau}) as a function of the rescaled biasing parameter $s^{\rm con}/D_r$ (See Eq.~\ref{equ:s_con}).  
 {\bf(d)}~The data from panel (c) plotted using rescaled variables, showing data collapse.
\textit{Parameter values (all panels)}: $N=50$, $\phi=0.65$, $n_c = 10^3$, $t_{\mathrm{max}} = 10^3$.
}
\label{fig:ws-nu}
\end{figure}

Recall that the persistence length $\lp$ measures the strength of the active self-propulsion, compared to passive diffusion.
Fig.~\ref{fig:ws-nu} shows results obtained by cloning for $\lp\geq2$, where the active propulsion is significant.  Note that since we focus throughout on $s<0$, the point corresponding to the unbiased (natural) dynamics is {on the right} of the graphs and the strength of the bias {increases from right to left}.  As the bias becomes more negative, both the active work and the orientational order parameter increase slowly at first, before showing a more rapid increase.  Similar to (\ref{equ:I''}),
\beq
w'(s) =  -\lim_{\tau \to\infty} \tau N [ \langle w_\tau^2 \rangle_s - \langle w_\tau \rangle_s^2 ]
\label{equ:w'}
\eeq
so a rapid change in $w(s)$ corresponds to a large variance in the biased ensemble.
The analogy with thermodynamic phase transitions suggests that the critical point $s=-s^*$ coincides with the point where $|w'(s)|$ is maximal.
Panels (b,d) of Fig.~\ref{fig:ws-nu}  show that appropriate changes of variable can be used to collapse the data for different $\lp$, details are given just below, in Section~\ref{sec:enhance}.

\begin{figure}
\centering
\begin{overpic}[trim=0 0 0 0, width=0.48\textwidth]{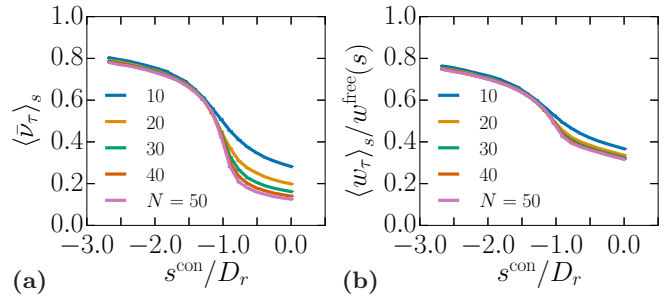}
\put (0, 0) {\texttt{\bf (a)}}
\put (\xp, 0) {\texttt{\bf (b)}}
\end{overpic}
\caption{Biased averages of the active work $\langle w \rangle_s$ [panel (a)] and of the polarisation norm $\langle |\bm{\nu}| \rangle_s$ [panel (b)] as functions of the rescaled biasing parameter $s^{\rm con}/D_r$ (See Eq.~\ref{equ:s_con}) for number of particles $N = 10, 20, 30, 40, 50$. \textit{Parameter values}: $\lp=5$, $\phi=0.65$, $n_c = 10^3$, $t_{\mathrm{max}} = 10^3$.}
\label{fig:fss}
\end{figure}

Fig.~\ref{fig:fss} {shows} the dependence on system size.
For $-s^*<s<0$, the order parameter decreases with $N$ as in (\ref{equ:para}), while for $s<-s^*$ it depends weakly on $N$ as in (\ref{equ:ferro}).  Together with the large variance (\ref{equ:w'}), this justifies our identification of $s=-s^*$ as a critical point at which symmetry is spontaneously broken. The dependence of $w(s)$ on $N$ is weaker, this function should be continuous at a critical point, with a singularity in its derivative at $-s^*$; this is consistent with the data.

\subsection{Enhancement of self-propulsion in biased ensembles}
\label{sec:enhance}

We now discuss the reasons for the data collapse observed in
Fig.~\ref{fig:ws-nu}(b,d). 
Note that there are two non-trivial contributions to $w_\tau$ in (\ref{equ:wdecomp}), which affect the biased ensemble (\ref{equ:ave-bias}) in different ways.
To separate their effects, observe that the average (\ref{equ:ave-bias}) in the biased ensemble may be reformulated as
\beq
\left\langle\mathcal{A}\right\rangle_s = \frac{\left\langle\mathcal{A} \, e^{-s N \tau w_{f,\tau} }\right\rangle_{v^{\rm con}_s} }
{\left\langle e^{-s N \tau w_{f,\tau}  } \right\rangle_{v^{\rm con}_s} }
\label{equ:ave-bias-vcon}
\eeq
where $w_{f,\tau}$ was defined in (\ref{equ:wf}); the averages on the right hand side are computed for the natural dynamics of a controlled ABP system in which the velocity $v_0$ in (\ref{equ:dr}) is replaced by
\beq
v^{\rm con}_s = v_0\left( 1 - \frac{2sD}{v_0^2} \right) \; .
\label{equ:vcon-s}
\eeq
This result was noted previously in~\cite{Nemoto2019,Grandpre2020}.
To highlight connections between optimal-control theory and large deviation theory~\cite{Touchette2015,Jack2015,Jack2020}, we refer generically to systems with modified equations of motion as \textit{controlled systems}, see also Sec.~\ref{sec:control}.
In this case, the only modification is the change in self-propulsion velocity, from $v_0$ to $v^{\rm con}$.

Eq.~(\ref{equ:ave-bias-vcon}) is an exact equality, even for finite $\tau$.  This is explained in Appendix~\ref{app:abp-LD}, by considering the biased time-evolution operators corresponding to (\ref{equ:ave-bias},\ref{equ:ave-bias-vcon}).
Comparing (\ref{equ:ave-bias},\ref{equ:ave-bias-vcon}), the equations of motion of the system have been modified, and the contribution $w_{\eta,\tau}$ has been removed from the exponential biasing factor.

Non-interacting ABPs have $w_{f,\tau}=0$, in which case this construction allows a full solution of the large-deviation problem.
The average (\ref{equ:ave-bias-vcon}) in the biased ensemble reduces to an unbiased steady-state average for ABPs where only the
self-propulsion velocity is modified, leading to an active work
\beq
w^{\rm free}(s) 
= 1 - \frac{2 s D}{v_0^2}.
\label{equ:w-free}
\eeq
The linearity of this function means that the SCGF and the rate function are quadratic: there are no collisions so the only fluctuations of $w_\tau$ come from the (Gaussian) noise, via \eqref{equ:weta}.

We now consider the effect of of interactions,  in the modified system of \eqref{equ:ave-bias-vcon}, with propulsion velocity $v^{\rm con}$.
The normalised work $w_{f,\tau}$ is not a particularly natural quantity in the modified system, because it contains a normalisation factor $v_0$ from (\ref{equ:wf}).
It is more natural to rescale $w_f$ by $w^{\rm free}$: the quantity that appears in the exponent of (\ref{equ:ave-bias-vcon}) is then
\beq
s w_{f,\tau} =s^{\rm con} \frac{{w}_{f,\tau}}{w_{\rm free}}
\eeq
with
\beq
s^{\rm con}  = s \left( 1 - \frac{2sD}{v_0^2} \right) \; .
\label{equ:s_con}
\eeq
The field $s^{\rm con}$ is conjugate to the normalised work $({w}_{f,\tau}/w_{\rm free})$, it is the natural biasing parameter for the controlled system.
Both $s$ and $s^{\rm con}$ have dimensions of inverse time.

The data collapse in Fig.~\ref{fig:ws-nu}(b,d) is obtained by plotting the normalised work $\langle w_{f,\tau} \rangle_s/w^{\rm free}(s)$ against its conjugate variable $s^{\rm con}$, rescaled by $D_r$.  This shows that the biased ensemble (\ref{equ:ave-bias-vcon}) is controlled primarily by the dimensionless combination $s^{\rm con}/D_r$, with a much weaker dependence on $\lp$, at least in this regime where the active self-propulsion is strong.
Hence the decomposition of $w_\tau$ as (\ref{equ:wdecomp}) allows its large deviations to be analysed as a combination of two factors: the bias acts on $w_{\eta,\tau}$ to increase the self-propulsion; it acts on $w_{f,\tau}$ to generate alignment and collective motion.
The reasons why $s^{\rm con}$ should be scaled by $D_r$ and not by some other rate (for example $v_0/\sigma$) will be discussed in later Sections.

\subsection{The case $\lp=1$}
\label{sec:lp1}

\begin{figure}
\centering
\begin{overpic}[trim=0 0 0 0, width=0.48\textwidth]{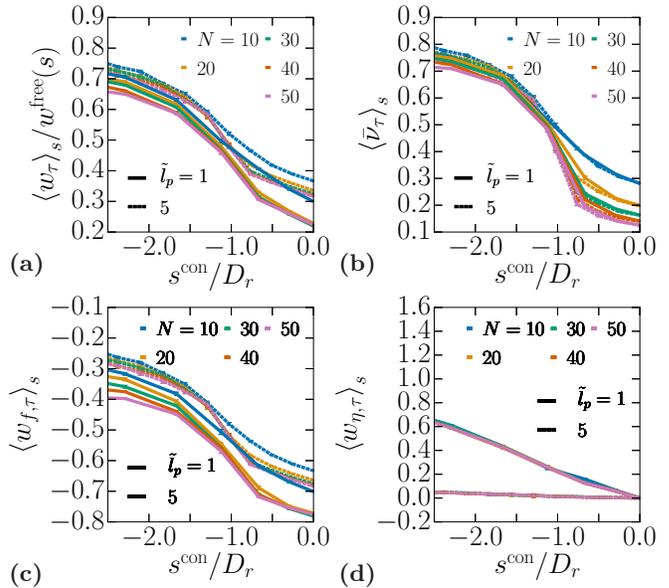}
\put (0, \yp) {\texttt{\bf (a)}}
\put (\xp, \yp) {\texttt{\bf (b)}}
\put (0, 0) {\texttt{\bf (c)}}
\put (\xp, 0) {\texttt{\bf (d)}}
\end{overpic}
\caption{{\bf (a)} Biased average of the active work rescaled by the biased free particle active work (See Eq.~\ref{equ:w-free}). {\bf (b)} Biased average of the polarisation norm $\langle \bar{\nu}_{\tau} \rangle_s$ (See Eq.~\ref{equ:bar-nu-tau}). {\bf (c)} Biased average of the force part of the active work. {\bf (d)} Biased average of the noise part of the active work. \textit{Parameter values}: $\phi=0.65$, $n_c=10^2$, $t_{\rm max}=10^2$.}
\label{fig:lp1}
\end{figure}

We now turn to cases where the self-propulsion is weaker, corresponding to smaller $\lp$.  Fig.~\ref{fig:lp1} shows results for several system sizes, comparing $\lp=1$ with $\lp=5$.  The order parameter $\langle \nubar \rangle_s$ no longer collapses perfectly but the qualitative behaviour is the same as for larger $\lp$. In particular the CM transition is robust.   However, the active work $\langle w_\tau\rangle_s$ no longer collapses as a function of $s^{\rm con}/D_r$, there are significant deviations.  In other words, the dependence of (\ref{equ:ave-bias-vcon}) can no longer be captured by the single dimensionless parameter $s^{\rm con}/D_r$, but it depends also on $\lp$ when that parameter is (relatively) small.

Physically, this can be rationalised by considering two mechanisms for events with large $w_{f,\tau}$ --  in practice, these are events where  particles spend less time in contact.  The first mechanism is CM -- if particles all move with fixed speed $v^{\rm con}$ in the same direction, they never collide, as discussed in Sec.~\ref{sec:symm-break-intro}.  A second mechanism is isotropic -- particles move in random directions, but tend to avoid each other when they get close.  When $\lp$ is large then the self-propulsion velocity dominates particles' relative velocity, and the CM mechanism dominates.  This is the regime where the data collapses in Fig.~\ref{fig:ws-nu}.
We argue that the isotropic mechanism is becoming important for smaller $\lp$, at which point the response to the bias $s$ acquires a more complicated dependence on $s/D_r$ and $\lp$.

Note that the isotropic mechanism does not rely on active self-propulsion and can be relevant in passive systems, when considering large deviations of quantities like $w_{f,\tau}$.  On the other hand, the CM mechanism is inherent to active systems.  It is therefore not surprising that the isotropic mechanism becomes more important for small $\lp$, where the system is behaving more like a passive fluid.  The limit $\lp\to0$ would be an interesting direction for future study, our expectation is CM still appears for $s^{\rm con} \lesssim -D_r$, but a more detailed understanding of the isotropic mechanism would be required, in order to establish this.

In addition, Fig.~\ref{fig:lp1}(d) shows the the noise contribution to the active work $w_{\eta,\tau}$.   For $\lp=1$, this contribution to $w_\tau$ is comparable to the force part $w_{f,\tau}$ while for larger $\lp$, the force contribution $w_{f,\tau}$ is the larger contribution.  From (\ref{equ:w-free}), the noise part is of order $-sD/v_0^2$; taking $s\sim D_r$ then this scales as $\lp^{-2}$ and is indeed small in the active limit of large $\lp$.  The force part $w_{f,\tau}$ is of order unity in this limit.

\section{Spontaneous alignment of two RTPs}
\label{sec:two-rtps}
\begin{figure}
\centering
\begin{overpic}[trim=0 0 0 0, width=0.48\textwidth]{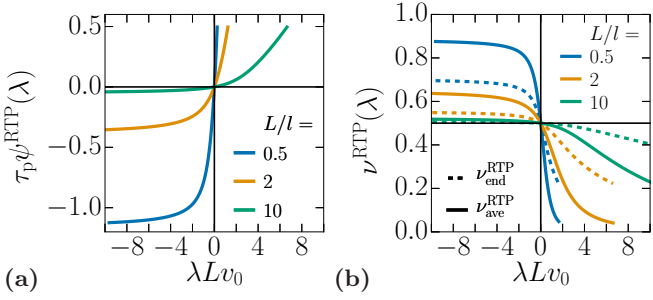}
\put (0, 0) {\texttt{\bf (a)}}
\put (\xp, 0) {\texttt{\bf (b)}}
\end{overpic}
\caption{{\bf (a)} Rescaled CGF $\tilde{\psi}^{\rm RTP} = \rtptime\psi^{\rm RTP}$ from (\ref{equ:rtp-lambda-psi-exact}) as a function of the rescaled biasing parameter $\tilde{\lambda} = \lambda l v_0$, for different values of the rescaled ring length $\tilde{L} = L/l$. {\bf (b)} Polarisation $\nu^{\rm RTP}_{\rm ave}$ as defined in \eqref{equ:nu-rtp-ave-nubar}, showing particle alignment for $\tilde\lambda<0$.  The dashed lines show $\nu^{\rm RTP}_{\rm end}$, which is evaluated at the end of the trajectory instead of as an average over all times, see also (\ref{equ:nu-rtp-ave-nuend},\ref{equ:nu-rtp-end-exact}).
}
\label{fig:rtp}
\end{figure}

As a preliminary step before describing a detailed theory of CM,
we illustrate the mechanism for collective motion by an analytic computation of large deviations of the active work, for two RTPs on a one-dimensional periodic ring.
Since the system is finite, there {cannot be any} spontaneous symmetry breaking, but we do find that the particles tend to align their orientations when biased to large active work.
An outline of the large deviation calculations can be found in Appendix~\ref{app:rtp}. Analysis of the unbiased behaviour of the system can be found in Ref.~\cite{Slowman2016}.

The RTPs have positions $r_i$ for $i=1,2$, with periodic boundaries, in a domain of size $L$.  Their active self-propulsion velocities are $\alpha_i v_0$ with $\alpha_i=\pm 1$.
Particle $i$ tumbles with rate $\rtptime^{-1}$, which corresponds to $\alpha_i$ changing its sign, and we introduce $l = v_0 \rtptime$ the persistence length.  The particles interact by a pair potential $V$, and there is no thermal diffusion. Let the particle separation be $r_{12}=|r_1-r_2|$.  Hence the equation of motion (between tumbles) is
\beq
\dot{r}_i = \alpha_i v_0 - \frac{\partial}{\partial r_i} V(r_{12}) \; .
\label{equ:eom-rtp}
\eeq

The potential $V$ is short-ranged with $V(r_{12})\to\infty$ as $r_{12}\to0$, and a length scale $\epsilon$ such that $V(r_{12})=0$ for $r_{12}>\epsilon$.  For particles in contact, there is a particular distance $r^*$ (less than $\epsilon$) such that $V'(r^*)=-v_0$, so that two particles with opposite orientations can have a force-balanced state with $\dot{r}_i=0$.  We focus on the limit of hard particles such that $\epsilon\to0$ and also $r^*\to0$.

The unnormalised instantaneous rate of active work is defined analogous to (\ref{equ:wf})  as
\begin{align}
\label{equ:rtp-bias}
\dot{w}^{\rm RTP}_f
 = v_0 (\alpha_1 - \alpha_2) \frac{\partial}{\partial r_1} V(r_{12})\\
w^{\rm RTP}_f = \lim_{\tau \to \infty} \frac{1}{\tau} \int_0^{\tau} \dot{w}^{\rm RTP}_f(t) \, \mathrm{d}t
\end{align}
while there is no analogue of the noise term $w_\eta$ because there is no thermal noise.  In the limit of hard particles one has simply that
$
\dot{w}^{\rm RTP}_f = -2v_0^2
$
if the particles are in a force-balanced touching state [for example $r_{12}=0$ with $\alpha_1=1=-\alpha_2$]
and $\dot{w}^{\rm RTP}_f=0$ otherwise.
We denote by $\lambda$ the biasing field  conjugate to the active work, and introduce the cumulant generating function
\beq
\psi^{\rm RTP}(\lambda) = \lim_{\tau \to \infty} \frac{1}{\tau} \log \left\langle e^{-\lambda \int_0^{\tau} \dot{w}^{\rm RTP}_f(t) \, \mathrm{d}t} \right\rangle
\label{equ:rtp-cgf}
\eeq
such that $\langle w_f^{\rm RTP} \rangle_{\lambda} = - \partial_{\lambda} \psi^{\rm RTP}(\lambda)$ as usual.  Here $\lambda$ is playing the role of $s$ in the ABP system.

This quantity can be obtained by solving an eigenproblem, as discussed in Appendix~\ref{app:rtp}.
Fig.~\ref{fig:rtp}(a) shows that $\psi^{\rm RTP}$ converges to a constant when $\lambda \to -\infty$, therefore $\lim_{\lambda \to -\infty} \langle w_f^{\rm RTP} \rangle_{\lambda} = 0$, indicating that collisions are completely suppressed in this regime.

The analogue of \eqref{equ:bar-nu-tau} in this system is
\beq
\nubar^{\rm RTP}_\tau =  \frac{1}{\tau} \int_0^\tau \frac{ 1 + \alpha_1(t) \alpha_2(t)  }{2} \, \mathrm{d}t \, ,
\eeq
which is between $0$ and $1$, with an average value of $1/2$ in the unbiased state where the $\alpha$s are independent.
Its average value in the biased ensemble is
\beq
\label{equ:nu-rtp-ave-nubar}
\nu^{\rm RTP}_{\rm ave}(\lambda) = \lim_{\tau \to \infty}\langle \nubar^{\rm RTP}_\tau \rangle_\lambda.
\eeq
Note that this quantity is averaged over the whole trajectory, and its determination requires one to consider both the left- and right-eigenvectors of the associated eigenproblem.  The computation is described in Appendix~\ref{app:rtp}, which also considers the quantity
\beq
\label{equ:nu-rtp-ave-nuend}
\nu^{\rm RTP}_{\rm end}=\left\langle \frac{1 + \alpha_1(\tau) \alpha_2(\tau)}{2} \right\rangle_\lambda 
\eeq
which measures degree of alignment at the final time $\tau$.

Fig.~\ref{fig:rtp}(b) shows results.  Starting from the zero-bias state where all configurations are equiprobable ($\nu^{\rm RTP} = 1/2$), the polarisation increases as $\lambda$ is reduced from zero, corresponding to
large positive fluctuations of the active work.
For large negative $\lambda$,
the polarisation eventually reaches a plateau value. At fixed persistence length, the larger systems have weaker polarisation.
On the contrary, the polarisation decreases for positive $\lambda$, indicating that anti-aligned states are more probable than aligned states, so particles spend more time in collision.  
As usual, the time-averaged measurement $\nu_{\rm ave}^{\rm RTP}$ responds more strongly to the bias than the corresponding measurement $\nu_{\rm end}^{\rm RTP}$ at the final time~\cite{Nemoto2016}.

The main conclusion from this analysis is that a one-dimensional system of two self-propelled particles already shows that biasing towards fewer collisions promotes the alignment of the particles' orientations. We also describe in Appendix~\ref{app:rtp-scaling} a scaling regime that is relevant when the system is very large, which allows some simplification of the resulting expressions.

\section{Large deviation mechanism}
\label{sec:mechanism}

We now present an analysis of the mechanism by which CM occurs in the system of many interacting ABPs, as in Figs.~\ref{fig:rate}-\ref{fig:lp1}.  For two RTPs, we have seen that alignment is a natural mechanism for suppressing collisions between particles.  The same is true for ABPs.  To characterise the CM state we compare the biased ensemble (\ref{equ:ave-bias}) with other ensembles that we define either by modifying the equations of motion of the system (via control forces), or by applying different kinds of bias to the system.

\subsection{Control forces}
\label{sec:control}

Within large deviation theory, it is often useful to compare biased ensembles like (\ref{equ:ave-bias}) with ensembles where the dynamics of the system is modified by control forces~\cite{Touchette2015, Jack2015, Jack2020}.  In principle, biased ensembles can be reproduced exactly by a suitable (optimally-controlled) system.  However, these optimal control forces cannot usually be derived exactly, except in idealised models.

As a generic controlled ABP system we take equations of motion
\begin{align}
\dot{\bm{r}}_i & =  v^{\rm con} \bm{u}_i
- D  \nabla_i ( U + U^{\rm con} ) + \sqrt{2D} \, \bm{\eta}_i \;,
\nonumber\\ \dot{\theta}_i & = - D_r \frac{\partial  U^{\rm con}}{\partial \theta_i}  + \sqrt{2 D_r} \xi_i,
\label{equ:dr-con-gen}
\end{align}
where $U^{\rm con}$ is a control potential (dependent on all particle positions and orientations)
and $v^{\rm con}$ is a parameter (independent of positions and orientations).

For ABPs in the biased ensemble (\ref{equ:ave-bias}), we explain in Appendix~\ref{app:abp-LD} that the optimally-controlled system has
\begin{align}
v^{\rm con} & =v^{\rm con}_s \nonumber \\
U^{\rm con} & = U_s^{\rm opt} \; ,
\label{equ:con-opt}
\end{align}
where $v^{\rm con}_s$ was defined in (\ref{equ:vcon-s}) and $U_s^{\rm opt}$ must be determined from the solution to an eigenproblem, see
(\ref{equ:eigen-abp2},\ref{equ:U-opt-F}).  Note that the equations of motion for the optimally-controlled process have exactly the same random noise terms as the original process (\ref{equ:dr}), this is a general feature of large deviations in this class of system~\cite{Chetrite2015}.

We consider several controlled systems where the orientational dynamics of ABPs is modified by long-ranged coupling that favours particle alignment.  We test numerically how well they capture the properties of the biased ensemble (\ref{equ:ave-bias}), and hence the true large deviation mechanism.  This is achieved by deriving bounds on the rate function $I(w)$.  The bounds would be exact equalities if our approximations for the optimally controlled dynamics were exact.  In fact the bounds are close but not exact -- from this we conclude that our approximations capture important features of the large deviation mechanism, but they also miss important parts of the physics. In contrast to this work, the bound considered in~\cite{Grandpre2020} is restricted to $U^{\rm con}=0$, in which case the control forces only affect the self-propulsion velocity $v^{\rm con}$.

To construct bounds, let $P$ be the path probability measure for the ABPs, and let $P^{\rm con}$ be the corresponding measure for the system with control forces.  Averages in the controlled system are denoted $\langle \ldots \rangle_{\rm con}$.  Then for any (ergodic) controlled system with $\langle w_\tau \rangle_{\rm con} = w$ we have
\beq
I(w) \leq \lim_{\tau\to\infty} \frac{1}{N\tau} {\cal D}_{\rm KL}(P^{\rm con}||P)
\label{equ:I-bound-gen}
\eeq
where ${\cal D}_{\rm KL}(Q||P)$ is the
Kullback-Leibler (KL) divergence between distributions $P$ and $Q$ (see Eq.~\ref{equ:DKL-def}).
In the analogy between large deviation theory and thermodynamics,
(\ref{equ:I-bound-gen}) corresponds to the Gibbs-Bogoliubov inequality~\cite{Jack2020}.

As a general controlled system we take (\ref{equ:dr-con-gen}).
In this case the KL divergence can be computed, see (\ref{equ:DKL-S},\ref{equ:dS-con}). 
The key point is that if the control forces are optimal then (\ref{equ:I-bound-gen}) is an equality.
We denote the path probability distribution for this optimally-controlled system by $P^{\rm opt}_s$ so
\beq
I(w(s)) = \lim_{\tau\to\infty} \frac{1}{N\tau} {\cal D}_{\rm KL}(P^{\rm opt}_s||P) \; .
\label{equ:I-bound-opt}
\eeq
Averages with respect to $P^{\rm opt}_s$ also match averages in the biased ensemble, that is  $\langle {\cal A} \rangle_s \simeq \langle {\cal A} \rangle_{\rm opt}$ (see Eq.~\eqref{equ:ave-bias}), as $\tau\to\infty$.

As a general rule, the closer is the controlled system to the true fluctuation mechanism, the more accurate will be the bound (\ref{equ:I-bound-gen}).  The intuition is that choosing a controlled process corresponds to proposing a mechanism for the rare event, and this mechanism can occur with a particular probability of order $\exp[- {\cal D}_{\rm KL}(P^{\rm con}||P)]$. Hence, smaller values of ${\cal D}_{\rm KL}$ correspond to mechanisms that are exponentially more likely, and the mechanism that minimises ${\cal D}_{\rm KL}$ is an accurate representation of the rare event.

\subsection{Coupling among ABP orientations and upper bound on rate function}
\label{sec:control-argument}

As already discussed in~\cite{Nemoto2019}, the spontaneous breaking of symmetry for $s<0$ leads to a natural comparison with systems where torques act on the ABPs, so that their orientations tend to align.
This phenomenon is reminiscent of the flocking observed in systems of Vicsek particles \cite{Vicsek1995}, when the orientational coupling between neighbouring particles exceeds a threshold set by the effective temperature of the rotational dynamics.
To explore this effect, take as control potential an infinite-ranged (mean-field) coupling among the orientations, with strength $g>0$  corresponding to a ferromagnetic interaction.  That is,
\beq
U^{\rm con}_g = - \frac{gN}{D_r} |\bm{\nu}|^2
\label{equ:u-con-g}
\eeq
independent of particle positions.
The direction of the order parameter is determined by an angle $\varphi$ through $\bm{\nu} = |\bm{\nu}| ( \cos\varphi,\sin\varphi)$.
The equation for the ABP orientation in the controlled system can then be written as 
\beq
\dot\theta_{i} = - g |\bm{\nu}|\sin(\theta_i-\varphi)  + \sqrt{2D_r}\,\xi_i.
\label{equ:theta-con}
\eeq
Similar to the original ABPs, this orientational equation of motion is independent of all particle positions.  Hence, integrating out the particle positions leads to a mean-field system of interacting rotors, which is fully described by (\ref{equ:theta-con}).
For large $N$, appendix~\ref{app:MFrotors} shows that this system spontaneously breaks rotational symmetry at $g=D_r$. That is, for $g>D_r$ then $\langle |\bm{\nu}|\rangle_{\rm con}=O(1)$ as $N\to\infty$, but $\langle |\bm{\nu}|\rangle_{\rm con}=O(N^{-1/2})$ for $g<D_r$.  The resulting situation is similar to (\ref{equ:ferro},\ref{equ:para}).

It was argued in~\cite{Nemoto2019} that this very simple controlled model can already capture quite accurately the collective motion phase, and it can predict the rate function in this regime.  To explore this idea in more detail,
we write $P^{\rm con}_g$ for the path probability distribution for the ABP dynamics with control potential $U^{\rm con}_g$.  Using this distribution with Eq.~\ref{equ:I-bound-gen} yields an upper bound
\beq
I(w^{\rm con}_g) \leq \lim_{\tau\to\infty} \frac{1}{N\tau} {\cal D}_{\rm KL}(P^{\rm con}_{g}||P)
\label{equ:I-bound-upper}
\eeq
where $w^{\rm con}_g = \langle w_\tau \rangle_{\rm con}$ with control potential $U^{\rm con}_g$.
This bound would be an equality if the controlled model fully captured the CM phase.
It will be compared with the exact result in Sec.~\ref{sec:num-bounds}.

\begin{figure*}
\centering
\begin{overpic}[trim=0 0 0 0, width=0.96\textwidth]{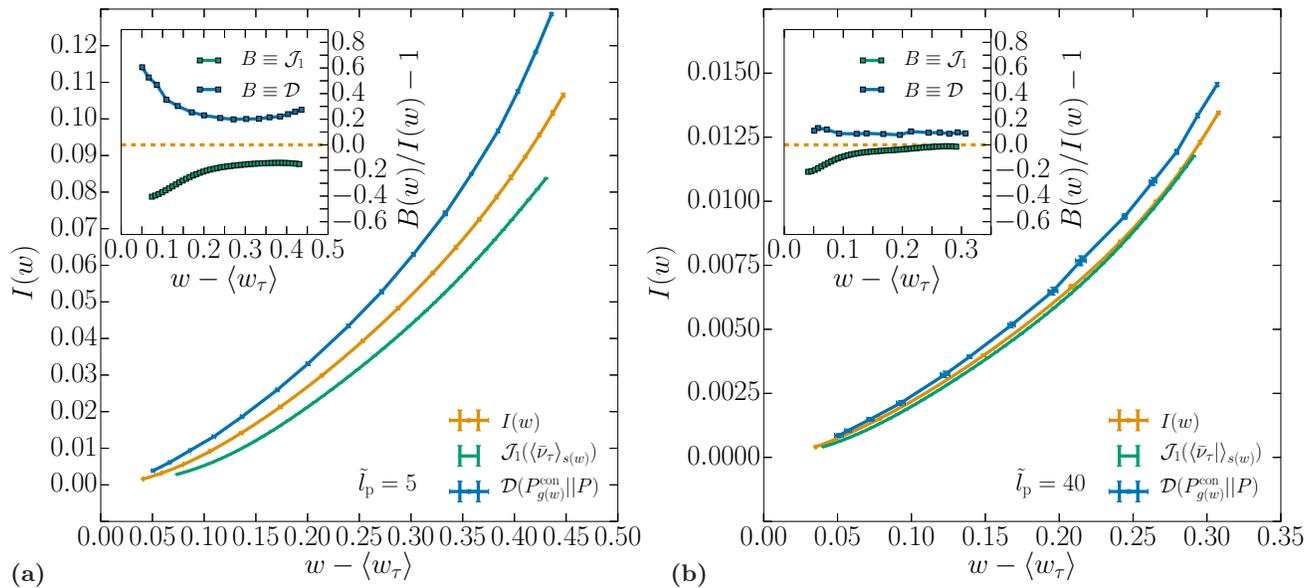}
\put (0, 0) {\texttt{\bf (a)}}
\put (\xp, 0) {\texttt{\bf (b)}}
\end{overpic}
\caption{
{\bf (a)} Rate function $I(w)$ from the cloning algorithm at $\lp=5$, compared with the upper and lower bounds (\ref{equ:I-bound-upper},\ref{equ:I-bound-lower}).
For the rate function, the solid line is a spline interpolation.  To evaluate bounds, results for $\langle \bar{\nu}_{\tau} \rangle_{s(w)}$ and $\mathcal{J}_1(\bar{\nu})$ are obtained by the cloning algorithm, and $\mathcal{D}(P^{\mathrm{con}}_{g(w)} || P)$ is obtained by simulation of the controlled dynamics.  Fourth order polynomial interpolation is then used to obtain values at the desired values of $w$.
{\bf (inset)} Relative error between the bounds and the rate function. 
{\bf (b)} Similar results as (a), for $\lp=40$.
\textit{General parameter values}: $N=10$, $\phi=0.65$. 
\textit{Cloning parameter values}: $n_c=10^3, t_{\mathrm{max}}=10^3, N_{\mathrm{runs}}=10$.
\textit{Parameters for controlled dynamics}: $t_{\mathrm{max}}=10^4$, $N_{\mathrm{runs}}=10$.
}
\label{fig:boundRate}
\end{figure*}

\subsection{Large deviations of order parameter and lower bound on rate function}

We also derive a lower bound on the rate function, similar to~\cite{Nemoto2019}.  To achieve this, we consider large deviations of the orientational order parameter.
Specifically we consider $\overline{\nu}_\tau$, defined in (\ref{equ:bar-nu-tau}) as the time-average of the modulus of the order parameter.  The statistical properties of other similar quantities (for example the modulus of the time average) have different dependence on  $N,\tau$, so some care is required in the following arguments.

The quantity $\overline{\nu}_\tau$ obeys an LDP as $\tau\to\infty$ which we write as
\beq
p(\nubar_\tau) \asymp \exp\left[- \tau N {\cal J}_1(\overline{\nu}_\tau)\right] \; .
\label{equ:LDP-nubar}
\eeq
where ${\cal J}_1$ is the rate function.  For large $N$, the function ${\cal J}_1$ can be obtained analytically, see Appendix~\ref{app:largedev-JJ}.
In particular, one has for large $N$ and small $\nubar$ that
\beq
\lim_{N\to\infty} {\cal J}_1(\overline{\nu}) = \frac{1}{2} D_r \nubar^2 + O(\nubar^4).
\eeq
Moreover, in this joint limit of large $N$ and small $\nubar$, the optimally controlled dynamics associated with these large deviations can be captured exactly by (\ref{equ:theta-con}).
The large deviations of $\nubar$ also satisfy a (general) bound analogous to (\ref{equ:I-bound-gen}), which is
\beq
{\cal J}_1(\nubar) \leq \lim_{\tau\to\infty} \frac{1}{N\tau}  {\cal D}_{\rm KL}(P^{\rm con}||P)
\label{equ:J1-bound}
\eeq
which holds for any controlled dynamics with $\langle |\bm{\nu}|\rangle_{\rm con} = \nubar$.

These results can be used to obtain a bound on $I(w)$.  We write $\nu(s) = \langle |\bm{\nu}| \rangle_s$.
Now consider (\ref{equ:J1-bound}) with $P^{\rm con}=P^{\rm opt}_{s}$,
as the optimally-controlled dynamics for large deviations of the active work, as in (\ref{equ:I-bound-opt}).
This optimally-controlled dynamics has  $\langle w_\tau\rangle_{\rm con} = w(s)$  and $\langle |\bm{\nu}|\rangle_{\rm con} = \nu(s)$.
Combining (\ref{equ:J1-bound}) with (\ref{equ:I-bound-opt}), we obtain
\beq
I(w(s)) \geq {\cal J}_1(\nu(s)) \; .
\label{equ:I-bound-lower}
\eeq
This is a lower bound on $I(w)$, which was derived in~\cite{Nemoto2019} by the contraction principle of large deviations.
From (\ref{equ:J1-bound}), the bound (\ref{equ:I-bound-lower}) is exact if the optimally controlled dynamics for large deviations of $w$ (that is {$P^{\rm opt}_s$}) is also an optimally-controlled dynamics for large deviations of $\nubar$.

Note that evaluation of (\ref{equ:I-bound-lower}) requires knowledge of $\nu(s)$, which is not available analytically -- instead one must perform cloning simulations.  For this reason, (\ref{equ:I-bound-lower}) is not a predictive result.  However, it is a useful result because the accuracy of the bound reveals the extent to which the (unknown) mechanism for large deviations of $w$ is similar to the (known) mechanism for large deviations of $\nubar$, as we now discuss.

\subsection{Numerical evaluation of bounds}
\label{sec:num-bounds}

Fig.~\ref{fig:boundRate} shows results for the rate function $I(w)$ (Sec.~\ref{sec:large-deviations}), compared with the upper and lower bounds~(\ref{equ:I-bound-upper},\ref{equ:I-bound-lower}).  We show data for $\lp=5$ as well as $\lp=40$, which was the case considered in~\cite{Nemoto2019}.

To evaluate the upper bound, we perform unbiased simulations of ABPs with the rotational equation of motion given by Eq.~\ref{equ:theta-con}, over a range of torque parameters $g$.  We compute ${\cal D}_{\rm KL}(P^{\rm con}_{g}||P)$ using (\ref{equ:DKL-con-g}) and the average active work $\langle w_\tau \rangle_{\rm con}$.  The upper bounds in Fig.~\ref{fig:boundRate} are parametric plots using these data (with $g$ as the parameter).

For the lower bound, we compute $\mathcal{J}_1(\bar{\nu})$ from cloning simulations of rotors (see Appendix~\ref{app:nu1}) and $\nu(s)$ from cloning simulations of ABPs, then compose these functions to obtain the right hand side of Eq.~\ref{equ:I-bound-lower}. We stress that the number of particles $N$ and the rotational diffusivity $D_r$ have to be consistent between simulations for this comparison to hold.

The bounds of Fig.~\ref{fig:boundRate} capture the main features of the rate function but there are significant deviations, especially for smaller $\lp$.
We note in particular that (i) the bounds (\ref{equ:I-bound-upper},\ref{equ:I-bound-lower}) are not accurate for small values of $w-\langle w\rangle_0$ but become accurate for larger $w$; (ii) the bounds are more accurate for larger $\lp$.  We now discuss these observations.

\subsection{CM (symmetry-broken) state}

From Fig.~\ref{fig:ws-nu}, the system is in the collective motion phase for $w>w^*$ with $w^* \approx 0.4$, weakly dependent on $\lp$.
For $\lp=5$, the CM phase is $w-\langle w_\tau\rangle \gtrsim 0.15$.  In this range, the inset of Fig.~\ref{fig:boundRate} shows that upper and lower bounds are accurate to around 20\% of the rate function.   For $\lp=40$, the accuracy of the lower bound is even better.

In both cases, the lower bound is more accurate.  Part of this effect may be attributed to the fact that the upper bound does not account for the enhancement of self-propulsion in the biased ensemble (Sec.~\ref{sec:enhance}).  The lower bound does account (at least partly) for this effect, since it uses the functions $w(s)$ and $\nu(s)$ as computed in cloning simulations.  The upper bound could be improved by including the controlled velocity $v^{\rm con}$ as an additional variational parameter in (\ref{equ:I-bound-gen}) and optimising it numerically, similar to~\cite{Jacobson2019,Grandpre2020}. However, it is sufficient for our argument to keep $v^{\rm con}=v_0$.

The conclusion for this regime is that fluctuations of the active work are strongly coupled to those of the orientational order parameter.  As a result, the bounds (\ref{equ:I-bound-upper},\ref{equ:I-bound-lower}) can capture the behaviour of the rate function almost quantitatively.  As noted in~\cite{Nemoto2019}, the log-probability of a large fluctuation of $w_\tau$ is almost the same as that of the corresponding orientational fluctuation.

\subsection{Isotropic state $\langle w_\tau \rangle < w < w^*$}

When the active work is close to its average value, the system does not break symmetry (recall Fig.~\ref{fig:ws-nu}) and one also
sees that the bounds in Fig.~\ref{fig:boundRate} do not capture the rate function in an accurate way.  The symmetry-breaking transition happens at $w=w^*$.  As $N\to\infty$, isotropic systems (with $w<w^*$) have $\nu(s)=0$ so the lower bound (\ref{equ:I-bound-lower}) tends to zero.  The rate function is not zero so the bound is not at all accurate in this range.  The conclusion for this regime is that fluctuations where $w_\tau$ is enhanced can also occur by an alternative fluctuation mechanism (not CM), and that this mechanism is dominant in the isotropic phase.  As noted above, the modified self-propulsion $v^{\rm con}=v^{\rm con}_s$ is relevant in this regime, and particles may also be repelled from each other without any alignment, as discussed in Sec.~\ref{sec:lp1}.  We explain {in Sec.~\ref{sec:landau}} that hydrodynamic density fluctuations are also relevant.  In other words, several effects act to enhance the active work in this regime, there is no collective motion, and the bounds of Fig.~\ref{fig:boundRate} do not follow the rate function accurately.

\begin{figure}
\centering
\begin{overpic}[trim=0 0 0 0, width=0.48\textwidth]{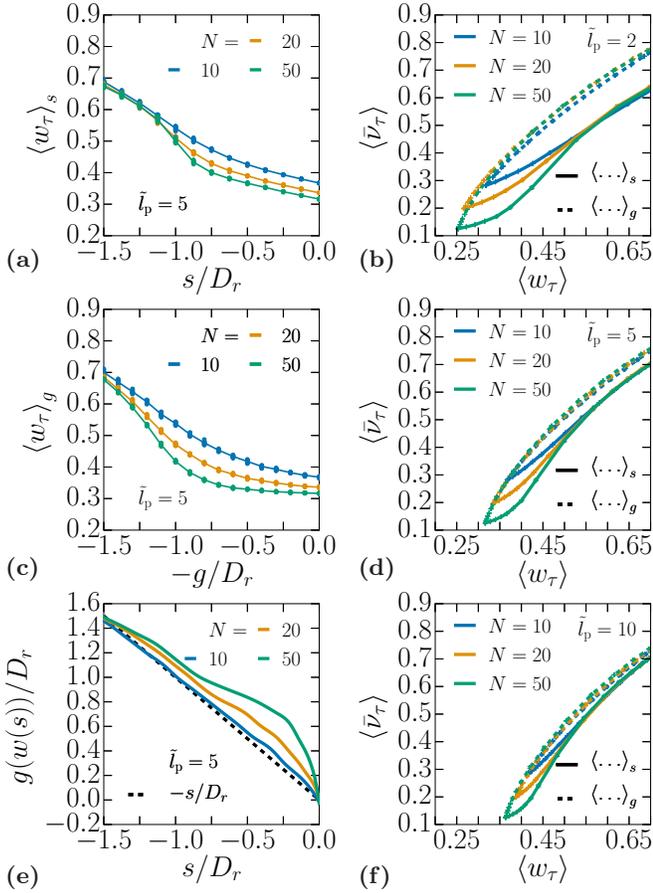}
\put (0, 70) {\texttt{\bf (a)}}
\put (40, 70) {\texttt{\bf (b)}}
\put (0, 35) {\texttt{\bf (c)}}
\put (40, 35) {\texttt{\bf (d)}}
\put (0, 0) {\texttt{\bf (e)}}
\put (40, 0) {\texttt{\bf (f)}}
\end{overpic}
\caption{{\bf (a)} Active work as a function of the biasing parameter from cloning simulations. Solid lines are spline interpolation.
{\bf (b)} Parametric plot of $\left<|\bm{\nu}|\right> = \left<\frac{1}{\tau} \int_0^{\tau} \mathrm{d}t \, |\bm{\nu}(t)|\right>$ as a function of the active work.  We show results for biased ensembles ($\langle\ldots\rangle_s$) and for controlled dynamics ($\langle\ldots\rangle_g$).
{\bf (c)} Active work as a function of the torque parameter $w$ in the controlled (modified) dynamics. Solid lines correspond to spline interpolations, points correspond to numerical data.
{\bf (d)} Similar to (b), with $\lp=5$.
{\bf (e)} Composition of the relation $g(w)$ from modified dynamics and $w(s)$ from cloning.
{\bf (f)} Similar to (b,d), with $\lp=10$.
{\textit{General parameter values}: $\phi=0.65$. \textit{Cloning parameter values}: $N_{\mathrm{clones}}=10^3$, $t_{\mathrm{max}}=10^2$, $N_{\mathrm{runs}}=10$. \textit{Controlled dynamics parameter values}: $t_{\mathrm{max}}=10^4$, $N_{\mathrm{runs}}=10$.}}
\label{fig:torque_parameter}
\end{figure}

\subsection{Comparison with controlled system}

To conclude our discussion of orientational fluctuations, we consider Fig.~\ref{fig:torque_parameter} which further illustrates the relationship between controlled dynamics and the large-deviation mechanisms.
 Panels {(a,c)} show the dependence of the active work, as either $s$ or $g$ is varied.  Also,
panels (b, d, f) show the orientational order parameter.

Note that $s$ is a biasing field in the sense of large deviations while $g$ is physical coupling between ABP orientations -- nevertheless, the response to both kinds of perturbation is similar, in that both $\langle w_\tau\rangle_\lambda$ and $\langle |\bnu|\rangle_\lambda$ increase.  The responses are different when the system is close to its unbiased steady state, in that the biased ensemble responds mostly by a change in $\langle w_\tau\rangle_\lambda$ while the controlled system responds mostly by increasing $\langle|\bnu|\rangle_\lambda$.  However, when the system is perturbed further from its steady state, both systems respond by entering a CM state, in which their behaviour is similar, with long-ranged orientational order and enhanced active work.

Finally, Fig.~\ref{fig:torque_parameter}(e) compares how large a coupling $g$ is required in order to achieve an active work equal to $w(s)$.
In the CM phase, one sees that $g(w(s))\approx -s$.  From (\ref{equ:g>Dr}),  rotational symmetry is spontaneously broken for the controlled system with $g>D_r$.  Hence  Fig.~\ref{fig:torque_parameter}(b, d, f) are consistent with the observation from Fig.~\ref{fig:ws-nu} that symmetry is broken in the biased ensemble for $s<-s^*$ with $s^*\approx D_r$.  A concrete theoretical explanation for this effect will be given in the next Section.

\section{Landau-Ginzburg Theory}
\label{sec:landau}

We have explained that many aspects of the collective motion phase of ABPs in terms of mean-field interactions among their orientations.  In this section we discuss how this mean-field picture can be embedded in a hydrodynamic theory, similar to Macroscopic Fluctuation Theory~\cite{Bertini2015}.

\subsection{Theory}

The minimal description that allows modelling of the symmetry-broken state is to consider a local density field $\rho$ and a corresponding polarisation field $\bm{P}$.  These are defined on hydrodynamic length scales: given a mesoscopic region $\Omega_{\bm r}$ centred at $\bm{r}$, we define
\begin{align}
\rho(\bm{r})  &= \frac{1}{|\Omega_{\bm r}|} \int_{\Omega_{\bm r}} \sum_i \delta(\bm{r}-\bm{r}_i) {\rm d}\bm{r}
\label{equ:def-rho-field}
\\
\bm{P}(\bm{r})  &= \frac{1}{\rho(\bm{r})|\Omega_{\bm r}|} \int_{\Omega_{\bm r}} \sum_i \bm{u}_i \delta(\bm{r}-\bm{r}_i) {\rm d}\bm{r}
\end{align}
where $|\Omega_{\bm r}|$ denotes the volume of $\Omega_{\bm r}$. The polarisation $\bm{P}$ is normalised as the average orientation (so $|\bm{P}|^2 < 1$).

With this choice, it is notable that $\rho$ is a slow hydrodynamic field, in the sense that density fluctuations on large length scales $\ell$ relax on long time scales of order $\ell^{2}$.  On the other hand $\bm{P}$ is a fast field, in that polarisation fluctuations on all scales relax quickly (on a time scale of order $D_r^{-1}$) to quasi-steady states 
which depend in general on $\rho$.

As a minimal model for ABPs in biased ensembles, we propose (in a ``top-down'' coarse-grained approach) the following (It\=o) equations of motion for $(\rho,\bm{P})$, similar to~\cite{Cates2013}:
\begin{align}
\dot\rho & = -\nabla\cdot \bJ
\nonumber\\
\bJ &= \bJ_{\rm d} + \sqrt{2\sigma(\rho)} \bm{\eta} ,
\nonumber\\
\dot \bP &= - \gamma(\rho,\bP) \bF(\bP) + b(\rho,\bP)\nabla \rho + \sqrt{2\gamma(\rho,\bP)} \bm{\xi}
\label{equ:eom-field}
\end{align}
where $\sigma$ and $\gamma$ are noise strengths; $b$ is a coupling between polarisation and density gradients; $\bF$ is a thermodynamic force that acts on the polarisation; $\bJ_{\rm d}$ is the deterministic part of the current; and $\bm{\eta}$ is a Gaussian white noise with zero mean and variance $\langle \eta^{\alpha}(t) \eta^{\beta}(t^{\prime}) \rangle = \delta_{\alpha\beta} \delta(t - t^{\prime})$.  We take a deterministic current
\beq
\bJ_{\rm d} = v_0 \rho \bP -  D_{\rm c}(\rho) \nabla \rho
\label{equ:jd}
\eeq
where the first term incorporates the effect of self-propulsion, while $D_{\rm c}(\rho)$ is the hydrodynamic (collective) diffusion constant, which depends on density.
Our theory is restricted to states without MIPS, this requires that $D_{\rm c}(\rho)>0$ for all $\rho$. (In fact the diffusion constant is additionally renormalised by polarisation fluctuations, as discussed in Appendix~\ref{app:fluctu-hydr}, and it is the renormalised diffusion constant that should be positive.)
In the absence of MIPS,  it is consistent to assume that $\nabla\rho$ is $O(\ell^{-1})$ where $\ell$ is the hydrodynamic length scale. This is the reason that higher gradients of $\rho$ are neglected in (\ref{equ:eom-field}).

It is also useful to compare (\ref{equ:eom-field}) with~\cite{Kourbane2018}, which is a rigorous hydrodynamic theory for a similar model, except that the polarisation is a slow field in that case.  In the ABP context, their analysis corresponds to reducing $v_0$ and $D_r$ as the system size increases, so that the polarisation becomes a hydrodynamic variable.  This corresponds to an idealised limiting case, but it shows how such theories can be justified rigorously.  Comparing (\ref{equ:eom-field}) with the Toner-Tu theory~\cite{Toner1995} and other theories for CM such as~\cite{Farrell2012},  our minimal model (\ref{equ:eom-field}) has fewer couplings, in particular it lacks advective terms in the polarisation equation.  Such terms are not relevant at the level considered here, but might be required for a quantitative description of fluctuations in the CM phase.

 On the hydrodynamic scale, it is consistent to approximate the (total) active work of a trajectory as a function of the density and polarisation fields:
\beq
w_\tau N \tau = \int_0^{\tau} \int_{[0,L]^2} \omega(\rho, \bP)\, {\rm d}\bm{r} {\rm d}{t}
\label{equ:w-omega}
\eeq
where $\omega(\rho, \bP)$ is the typical (average) active work per unit volume, in a region with density $\rho$ and polarisation $\bP$.

Now consider a large system of
linear size $L$, and define hydrodynamic co-ordinates as
\beq
\tilde{\bm{r}}=\bm{r}/L \;, \qquad \tilde{t}=D t/L^2 \;.
\eeq
We also express the fields in these rescaled variables as $\tilde\rho(\tilde{\bm{r}},\tilde{t}) = \rho(\tilde{\bm{r}}L,\tilde t L^2 /D)$, and $\tilde\bP(\tilde{\bm{r}},\tilde{t}) = \bP(\tilde{\bm{r}}L,\tilde t L^2 /D)$. [Note, there is no rescaling of the fields themselves, hence $|\tilde\bP(\tilde{\bm{r}},\tilde{t})|^2\leq 1$ is the modulus of the local polarisation, and $\tilde\rho(\tilde{\bm{r}},\tilde{t})\sigma^2\pi/4$ is a local area fraction, of the same order as the global area fraction $\phi=\bar\rho\sigma^2\pi/4$.]
At the expense of some heavy notation, we consistently use tildes to indicate that the hydrodynamic rescaling has been performed.  Where derivatives act on fields with tildes, they are taken with respect to the rescaled variables, for example $\dot{\tilde\rho} = (\partial\tilde\rho/\partial \tilde t)$.  We define
$\tilde \bJ = \bJ L/D$ so that the continuity equation retains its form:
\beq
\dot{\tilde\rho} = - \nabla\cdot \tilde{\bJ} \; .
\label{equ:cont-tilde}
\eeq

Using (\ref{equ:eom-field}) and working in hydrodynamic co-ordinates, we obtain a formula for the probability of a trajectory in the biased ensemble,
\beq
\mathrm{Prob}\big[\tilde\rho,\tilde\bJ,\tilde\bP\big] \propto \exp\left( -L^2 \int_0^{\tilde\tau} \int_{[0,1]^2}
{\cal S}\big[\tilde\rho,\tilde\bJ,\tilde\bP\big] \, {\rm d}\tilde t {\rm d}\tilde x \right)
\label{equ:prob-field}
\eeq
which is valid only if (\ref{equ:cont-tilde}) holds (else the probability is zero); and the Lagrangian is
\beq
{\cal S} =  \frac{D}{4\sigma} |\tilde \bJ - \tilde{\bJ}_{\rm d}|^2
 +\frac{1}{4 D \gamma}  \left| \frac{D}{L}\dot{\tilde{\bP}} + \gamma \bm{f} L -b\nabla\tilde{\rho} \right|^2
+ \frac{s L^2}{D} \omega
\label{equ:lag}
\eeq
where $\tilde \bJ_{\rm d} = \bJ_{\rm d} L/D$ and we have omitted the dependence of $\sigma,\gamma,\omega$ on $(\tilde\rho,\tilde\bP)$, for ease of writing.

The terms in (\ref{equ:lag}) that involve $\bJ$ and $\omega$ are familiar from other analyses of large deviations in systems with hydrodynamic modes~\cite{Sollich2015, Dolezal2019}.  In particular, the fact that $s$ enters through the combination $sL^2$ is familiar from earlier studies~\cite{Appert2008}, it reflects the fact that hydrodynamic degrees of freedom respond strongly to the bias, because of their slow relaxation.   However, the factors of $L$ that appear in the term involving $\tilde{\bP}$ are not expected in hydrodynamic theories, they reflect the fact that $\bm{P}$ is a fast field.

\begin{figure}[t]
\centering
\begin{overpic}[trim=0 0 0 0, width=0.48\textwidth]{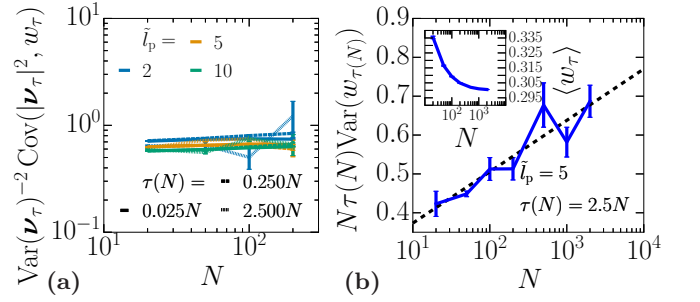}
\put (5, 0) {\texttt{\bf (a)}}
\put (\xp, 0) {\texttt{\bf (b)}}
\end{overpic}
\caption{{\bf (a)} Normalised covariance of the active work $w_{\tau}$ and the squared average polarisation $\boldsymbol{\nu}_{\tau}^2 = \left(\frac{1}{\tau} \int_0^{\tau} \boldsymbol{\nu}(t) \, \mathrm{d}t\right)^2$. {\bf (b)} Normalised variance of the active work with a $\log N$ fit in dashed line. \textit{Parameter values}: $\phi=0.65$, $N_{\mathrm{runs}} = $ {\bf (a)} $2 \cdot 10^2$ {\bf (b)} $6 \cdot 10^2$.}
\label{fig:variances}
\end{figure}

\subsection{Mean-field analysis}

We first consider the case where $(\tilde\rho,\tilde\bJ,\tilde\bP)$ are independent of space and time.
Hence we set $\tilde{\rho} = \bar\rho$ where
\beq
\bar\rho = \frac{N}{L^2}
\label{equ:rhobar}
\eeq
is the average density.  The action is minimised by taking $\tilde{\bJ} = (v_0L \bar\rho/D)\tilde{\bP}$, note that this is $O(L)$ when expressed in these hydrodynamic variables, because the self-propulsion leads to ballistic motion.
Then
\beq
{\cal S} = L^2 \left[ \frac{\gamma}{4 D}  \left| \bm{f}  \right|^2 + \frac{s  \omega(\bar\rho,\bP)}{D}  \right] \; .
\label{equ:S-mf}
\eeq

The next step is to use properties of ABPs to express the remaining quantities in terms of microscopic parameters.
Setting $s=0$ in (\ref{equ:S-mf}) and combining with (\ref{equ:prob-field}), one can read off the probability of a trajectory where the polarisation is fixed at $\bm{P}$ for all times between $0$ and $\tau$.  This same probability can be computed applying large-deviation theory to the microscopic ABP model: the relevant LDP is given in (\ref{equ:LDP-nu}), and Appendix~\ref{app:pol} explains how the rate function ${\cal J}$ can be computed as the Legendre transform of a Mathieu function, see also~\cite{Grandpre2018}.  Hence
\beq
{\cal S} = \frac{L^2}{D} \left[ \bar\rho {\cal J}(\bP) + s  \omega(\bar\rho,\bP) \right] \; .
\label{equ:S-J-om}
\eeq

Recalling that $\omega$ is defined as the active work per unit volume for a system with prescribed density and polarisation, we have $\omega(\bar\rho,\bm{0})=\bar\rho\langle w_\tau\rangle$.   For $\bP\neq0$, the consistent choice is to define $\omega$ by considering an ensemble of trajectories that is biased by the polarisation.  Details are given in Appendix~\ref{app:pol}, we define $\langle \cdot \rangle_{\bm h}$ as an average analogous to (\ref{equ:ave-bias}), but with the bias acting on the polarisation, see (\ref{equ:bias-h}).
Then
\begin{equation}
\omega(\bar{\rho}, \bm{P}) = \langle \bar\rho w_{\tau} \rangle_{{\bm h}(\bP)}
\label{equ:biased-omega-hydro}
\end{equation}
where ${\bm h}(\bP)$ is defined by $\langle \overline{\bnu}_\tau \rangle_{{\bm h}(\bP)}=\bP$.
For small $\bP$, we show in Appendix~\ref{app:bias} that
\beq
\omega(\bar{\rho}, \bm{P}) = \bar\rho \left[ \langle w_\tau\rangle + \frac{c_\omega}{2} | \bP |^2 + O( | \bP |^4 ) \right]
\label{equ:omega-c}
\eeq
\begin{equation}
c_\omega
= \frac{  \bar \rho  \tau^2L^4  D_r^2}{2}  \mathrm{Cov}\left( w_{\tau}, |\bm{\nu}_{\tau}|^2\right)
\label{equ:ompp}
\end{equation}
where $ \mathrm{Cov}(x,y) = \langle xy \rangle - \langle x \rangle \langle y \rangle $ is the covariance in the natural (unbiased) ABP dynamics.  Fig.~\ref{fig:variances}(a) shows that this covariance is positive. That is, trajectories with larger polarisation also tend to have larger active work.

To analyse spontaneous symmetry breaking, we require a Taylor expansion of  ${\cal S}$ for small $\bP$.
The behavior of ${\cal J}$ at small polarisation can be obtained exactly, the result is (\ref{equ:J-small-nu}), which implies that
\beq
 {\cal J}( \bm{P} ) = \frac12 D_r |\bm{P}|^2 + O(|\bm{P}|^4) \; .
\eeq
Hence (\ref{equ:S-J-om}) becomes
\beq
{\cal S} = \frac{\bar\rho  L^2}{D} \left[  s\langle w_\tau \rangle + \frac12 |\bP|^2 ( D_r + sc_\omega ) + O( | \bP |^4 )  \right] \; .
\label{equ:S-landau}
\eeq
Finally, minimising the action, we predict spontaneous symmetry breaking for $s<-s^*$ with $s^* = D_r/c_\omega$.
From Fig.~\ref{fig:variances}(a), one sees that $c_\omega$ depends weakly on $D_r$, so this is consistent with the observation of Sec.~\ref{sec:symm-breaking}, that $s^* \propto D_r$.

Physically $c_\omega>0$ reflects the fact that breaking symmetry reduces collisions between particles and increases the active work.  This effect is quadratic in $\bP$, and so is the rate function $\cal J$ associated with symmetry-breaking events.  Hence both contributions in (\ref{equ:S-landau}) have the same scaling with $\bP$ (and with $L$), leading to a critical point {$s^*$ (independent of $L$), for which} the coefficients of the two relevant terms balance each other.

Since this result was obtained by a top-down coarse-grained approach, one can expect that it should be quite generic, in that the same theory would be a natural description of other active particles like RTPs.  The key ingredient here is that the coefficient $c_\omega$ in (\ref{equ:omega-c}) should be positive, which means by (\ref{equ:ompp}) that a larger polarisation is correlated (in the unbiased steady state) with a larger active work.  In ABPs, there is a clear mechanism for this, that particles' relative velocities are reduced if their align, because $\bm{u}_i$ is a unit vector.  In other active systems, this would have to be checked on a case-by-case basis.

\begin{figure}
\centering
\begin{overpic}[trim=0 0 0 0, width=0.48\textwidth]{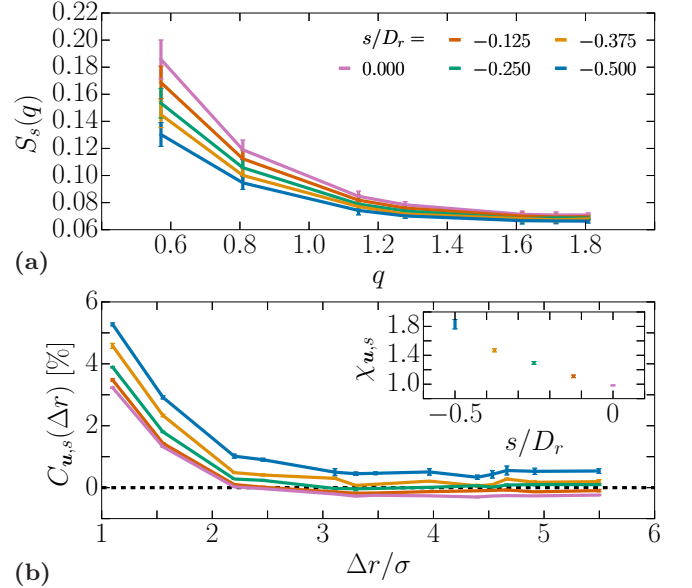}
\put (0, \yp) {\texttt{\bf (a)}}
\put (0, 0) {\texttt{\bf (b)}}
\end{overpic}
\caption{{\bf (a)} Structure factor as function of wavevector $k = 2\pi/\lambda$, zoomed on highest computed wavelengths, at biasing parameter $s$. {\bf (b, main plot)} Orientation correlation as a function of the distance at biasing parameter $s$. {\bf (b, inset)} Sum of the correlation function. \textit{Parameter values}: $N=10^2$, $l_p/\sigma=5$, $\phi=0.65$, $n_c=10^3$, $t_{\mathrm{max}}=10^2$.}
\label{fig:cuu}
\end{figure}

\subsection{Fluctuations}

This mean-field analysis predicts that symmetry is spontaneously broken for $s<-s^*$ with $s^*>0$.
This means that the system remains isotropic in a finite range around $s=0$.  However, the fact that $s$ enters (\ref{equ:lag}) as $sL^2$ indicates that the system can respond strongly to the bias already for very small $s$, via hydrodynamic fluctuations.
As explained in Appendix~\ref{app:fluctu-hydr}, the (fast) polarisation field is not relevant on the hydrodynamic scale so it is sufficient for small bias to consider a scalar theory for the density.  In this case density fluctuations and fluctuations of
the active work can be understood based on previous studies~\cite{Sollich2015, Dolezal2019}.

The behaviour depends on the sign of $\frac{\partial^2}{\partial \rho^2}\omega(\bar{\rho},\bm{0})$ which we abbreviate here by $\bar{\omega}^{\prime\prime}_0$.  For ABPs then $\bar{\omega}^{\prime\prime}_0<0$.
The first consequence of the hydrodynamic theory is that the variance of $w_\tau$ (under the natural dynamics) has a hydrodynamic contribution.
Appendix~\ref{app:fluctu-hydr} shows that the asymptotic variance as $L\to\infty$ behaves as
\beq
\bar\rho L^2 \lim_{\tau\to\infty} \tau \mathrm{Var}(w_\tau)_0 =
 \frac{(\bar{\omega}^{\prime\prime}_0 \sigma(\bar\rho))^2}{4\pi \bar\rho D_c(\bar{\rho})^3} [ \log L
 + O(1) ] \; .
 \label{equ:varW-logL}
 \eeq
Fig.~\ref{fig:variances}(b) shows numerical data that are consistent with this prediction.

The hydrodynamic theory of Appendix~\ref{app:fluctu-hydr}, also predicts for $\omega''_0<0$ that a system described by (\ref{equ:eom-field}) becomes inhomogeneous (phase separation) for
$s>s_c$ with~\cite{Sollich2015, Dolezal2019, Grandpre2020}
\beq
s_c L^2 = - \frac{2 \pi^2 D_c(\bar{\rho})^2}{\bar{\omega}^{\prime\prime}_0 \sigma(\bar{\rho})}.
\label{equ:sc}
\eeq
Recall that $\omega''_0<0$ which means that $s_c>0$.

For any negative value of $s$, one expects the system to be hyperuniform~\cite{Sollich2015,Dolezal2019}.
Write $\rho_{\bm q}$ for a Fourier component of the density field and $S_s(q) = \langle \rho_{\bm q}\rho_{-\bm{q}} \rangle_{s}$
for the structure factor in the biased ensemble, where $q=|\bm{q}|$.  The structure factor is derived in
(\ref{equ:Sq-s}) which shows that it behaves for small $q$ as
\beq
S_s(q) = \langle \rho_{\bm q}\rho_{-\bm{q}} \rangle_{s} \simeq
\begin{cases}
\chi_0
 , & s=0 ,
\\
b_s q , & s<0 . \end{cases}
\label{equ:Sq-HU}
\eeq
where $\chi_0=\sigma(\bar\rho) / D_c(\bar\rho)$ and $b_s$ are constants [the behavior of $b_s$ and can be read from (\ref{equ:Sq-s})].

Comparison of these theoretical predictions with numerical results requires some care because the numerical results are limited to small systems, while the hydrodynamic theory is valid only if the system size is much larger than all microscopic length scales.

The existence of inhomogeneous states for $s>s_c$ was already discussed in~\cite{Nemoto2019}.
For $s\leq 0$,
Fig.~\ref{fig:cuu}(a) shows the behaviour of the structure factor in the isotropic phase.  For $s=0$, the system is a homogeneous fluid so $\lim_{q\to0} S(q)$ must be a positive constant.  However, the numerical results show $S(q)$ increasing as $q$ is reduced towards zero.  The reason is that the system is not large enough to show the hydrodynamic behavior -- in a much larger system then smaller wavevectors would be accessible, and convergence of $S(q)$ to its small-$q$ limit would be apparent.  Similarly, for $s<0$, one expects in very large systems to observe $S(q) \propto q$ at small $q$, but this is not apparent from our numerical results in small systems.   Despite these restrictions, the numerical results show that density fluctuations are suppressed in biased ensemble with $s<0$, which is qualitatively consistent with the theory, and with the physical reasoning that reduced density fluctuations tend to suppress collisions and enhance the active work.

Density fluctuations also have consequences for the symmetry breaking transition at $s=-s^*$.  First, the transition takes place between a hyperuniform isotropic state and a symmetry-broken state (which may also be hyperuniform).   In this case the simple relationship (\ref{equ:w-omega}) should be adjusted, to account for the fact that the density fluctuations near $s^*$ are different from those of the biased ensemble (\ref{equ:bias-h}).  This effect presumably shifts the value of $s^*$ but we expect qualitative predictions of mean-field theory to remain valid.

An additional question is how density fluctuations couple with those of the polarisation, close to the critical point ($s=-s^*$).
If one ignores the coupling between $\bP$ and $\rho$ in (\ref{equ:lag}), one expects a transition in the universality class of an XY model in $(2+1)=3$ dimensions (two spatial dimensions and one of time).  This situation is familiar in quantum phase transitions for rotors~\cite{Sachdev1999}.

However, the coupling to a hydrodynamic density field may change this universality class.
For example, in the Vicsek model, a locally-aligning interaction leads to clustering of the particles. As a result, symmetry breaking for the orientations is coupled with collective motion and with phase separation~\cite{Vicsek1995, Chate2020}.
By contrast, for the phase transition considered here, clustering is suppressed (hyperuniformity), see Fig.~\ref{fig:cuu}(a).
Also, Fig.~\ref{fig:cuu} shows the spatial orientation correlation function
\beq
\begin{aligned}
C_{\bm{u}, s}(\Delta r) &= \frac{\int_0^\tau \left\langle \sum_{i, j = 1}^N \bm{u}(\theta_i(t)) \cdot \bm{u}(\theta_j(t)) \, \delta_{ij}(\Delta r, t) \right\rangle_{s} \, \mathrm{d}t}{\int_0^\tau \left\langle \sum_{i,j = 1}^N \delta_{ij}(\Delta r, t) \right\rangle_{s} \, \mathrm{d}t}\\
\delta_{ij}(\Delta r, t) &= \delta (|\bm{r}_i(t) - \bm{r}_j(t)| - \Delta r)
\end{aligned}
\eeq
This average is performed over trajectories  extracted from the cloning algorithm \cite{Nemoto2016}.
In these (small) systems, the main effect of the bias on fluctuations in the isotropic phase seems to be a constant (infinite-ranged) additive contribution to $C$, see Fig.~\ref{fig:cuu}(b).  This contribution is of order $1/N$ so that
\beq
\begin{aligned}
\chi_{\bm{u}, s} &= \int C_{\bm{u}, s}(r) \, 2\pi r \, \mathrm{d}r\\
&= \frac{1}{N} \left\langle \left(\sum_{i=1}^N \bm{u}(\theta_i(t))\right)^2 \right\rangle_{t, s}
\end{aligned}
\eeq
is of order unity -- and exactly equal to 1 in natural dynamics ($s=0$).  Such behaviour can be accounted for in mean-field theory -- one might expect non-trivial spatial structure to emerge in larger systems but analysis of such effects is beyond the scope of this work.

\section{Conclusion}
\label{sec:conclusion}

We have analysed large deviations with enhanced active work in an ABP system.  As found in~\cite{Nemoto2019}, this results in spontaneous breaking of rotational symmetry and collective motion.  We presented numerical evidence and theoretical arguments that this transition occurs at $s=-s^*$ with $s^* \simeq D_r$, at least when $D_r$ is small ($\lp\gg 1$).  This means that CM sets in above a threshold $w^*$ for the active work, with $w^*>\langle w_\tau\rangle$.  This resolves an open question from~\cite{Nemoto2019}, as to whether $w^*=\langle w_\tau\rangle$.

We have compared the behaviour of the CM state with that of a controlled system where the ABP orientations interact via an infinite-ranged (mean-field) coupling.  This controlled system captures the large deviations semi-quantitatively.
Based on a hydrodynamic theory, we have also explained that we expect hyperuniform behaviour for values of the active work between $\langle w_\tau\rangle$ and $w^*$.
We discussed the extent to which the predictions of this theory should be generic in systems of self-propelled particles.

Despite this progress, several questions remain open, including the nature of the critical point where the symmetry is broken.
A related point is whether a controlled system with a more complicated (distance-dependent) coupling of orientations would capture the CM phase more accurately.
Another point of comparison is collective motion in aligning particles, such as the Vicsek model and its variants~\cite{Vicsek1995, Farrell2012, Solon2015}.  In that case, breaking of rotational symmetry is often accompanied by phase separation into dense (polar) and more dilute (apolar) regions, which leads to a first-order transition to CM~\cite{Chate2020}.

For the systems considered here, the evidence [for example Fig.~\ref{fig:cuu}(a)] is that density fluctuations are suppressed in the CM phase, contrary to the enhanced density fluctuations that one might expect in systems with polar clusters.
However, these numerical results for small systems are not sufficient to settle the nature of the phase transition and its fluctuations.  A detailed analysis of symmetry breaking within the Landau-Ginzburg theory of Sec.~\ref{sec:landau} would be an interesting direction for future work.

\begin{acknowledgments}
The authors acknowledge insightful discussions with Takahiro Nemoto, Julien Tailleur, and Thibaut Arnoulx de Pirey.
This work was funded in
part by the European Research Council under the EU's
Horizon 2020 Programme, grant number 740269.
\'EF acknowledges support from an Oppenheimer Research Fellowship from the University of Cambridge,
and a Junior Research Fellowship from St Catharine's College. MEC is funded by the Royal Society.

All the codes that were developed and used in this project are made freely available under the MIT license \cite{GitHub}.
\end{acknowledgments}

\appendix

\section{Large deviations for ABPs}
\label{app:abp-LD}

\subsection{Eigenvalue problem}

Large deviations of $w_\tau$ can be analysed through the eigenvalues of an operator called the backwards generator.
For compactness of notation we define (only for this section)
$ 
\tilde{s}=(s/v_0)
$. 
Using results of \cite{Chetrite2015, Touchette2018}, the eigenvalue equation is
\begin{multline}
\psi(s) {\cal F}_s = D \sum_i (  \nabla_i -\tilde{s}  \bm{u}_i ) \cdot  (   \nabla_i -\tilde{s}  \bm{u}_i ) {\cal F}_s
\\ +  \sum_i  ( v_0 \bm{u}_i - D \nabla_i U ) \cdot (   \nabla_i -\tilde{s}  \bm{u}_i  ) {\cal F}_s
\\ + D_r \sum_i \frac{\partial^2 {\cal F}_s}{\partial \theta_i^2}
\label{equ:eigen-abp}
\end{multline}
where ${\cal F}_s={\cal F}_s(\bm{r}_1,\dots,\bm{r}_N,\theta_1,\dots,\theta_N)$ is the eigenvector and $\psi(s)$ the eigenvalue.
This may be simplified as
\begin{multline}
\psi(s) {\cal F}_s =
  \sum_i  [ (v_0 -2\tilde s D) \bm{u}_i - D \nabla_i U ] \cdot \nabla_i {\cal F}_s
\\ + \sum_i  \left[ D \nabla_i^2   {\cal F}_s+ D_r \frac{\partial^2  {\cal F}_s}{\partial \theta_i^2} \right]
\\ + \sum_i \left[  \tilde s  D (\bm{u}_i \cdot \nabla_i U)  + D\tilde s^2 -  \tilde{s}/v_0 \right] {\cal F}_s
\label{equ:eigen-abp2}
\end{multline}
where the first two lines correspond to the generator of an ABP system with a modified swim velocity $ v^{\rm con}_s$ as defined in (\ref{equ:vcon-s}), and the last line corresponds to a bias in which the only non-trivial term is $v_0 D (\bm{u}_i\cdot \nabla_i U)$, which is the scalar product between the swim velocity and the interparticle force, as it appears in the $w_f$ part of the active work.

The operator on the right hand side of (\ref{equ:eigen-abp2}) is equal (up to an additive constant) to the operator that generates the biased ensemble on the right hand side of (\ref{equ:ave-bias-vcon}).  The additive constant affects the eigenvalue but it does not affect the biased ensemble.  This is sufficient to establish (\ref{equ:ave-bias-vcon}).

From the solution to the eigenproblem, one may construct a corresponding optimally-controlled system, whose natural dynamics matches that of the biased ensemble.
The corresponding optimal control potential is
\beq
U^{\rm opt}_s= -2 \log {\cal F}_s \; .
\label{equ:U-opt-F}
\eeq

To construct the controlled system, consider the controlled dynamics (\ref{equ:dr-con-gen}).
The backwards generator for this model is ${\cal L}^{\rm con}$, which acts on functions ${\cal G}={\cal G}(\bm{r}_1,\dots,\bm{r}_N,\theta_1,\dots,\theta_N)$ as
\begin{multline}
{\cal L}^{\rm con}({\cal G}) = \sum_i \left[ D \nabla_i^2 {\cal G}
+ D_r \frac{\partial^2 {\cal G}}{\partial \theta_i^2} - D_r \frac{\partial U^{\rm con} }{\partial \theta_i} \frac{\partial {\cal G}}{\partial \theta_i} \right]
\\ + \sum_i \left[ v^{\rm con}\bm{u}_i - D  \nabla_i ( U + U^{\rm con} )  \right] \cdot \nabla_i {\cal G}
\end{multline}
Denoting the right hand side of (\ref{equ:eigen-abp2}) by ${\cal L}_s({\cal F}_s)$, the generator of the optimally-controlled dynamics may be then be derived using the general formula ${\cal L}^{\rm con}({\cal G}) = {\cal F}_s^{-1} {\cal L}_s({\cal F}_s{\cal G}) - \psi(s){\cal G}$, see for example~\cite{Chetrite2015}.
Combining these ingredients, the optimally-controlled dynamics is given by (\ref{equ:dr-con-gen},\ref{equ:con-opt}).

\subsection{KL divergence between controlled and natural dynamics}

To derive a formula for the KL divergence in (\ref{equ:I-bound-upper}), we recall its definition
\beq
{\cal D}_{\rm KL}(Q||P) = \int Q(x) \log\frac{Q(x)}{P(x)}  dx
\label{equ:DKL-def}
\eeq
where $Q,P$ are probability densities.
Using standard techniques in stochastic processes \cite{Onsager1953, Cugliandolo2017, Cugliandolo2019}, the path probability distribution for ABPs can be derived.  We work in Stratonovich calculus throughout.  The path probability is $P(X)\propto \exp[-\sum_i \int_0^\tau  {\cal S}_i(X(t)) dt ]$ where $X=\{\bm{r}_i, \theta_i\}_0^{\tau}$ indicates a path
and
\beq
{\cal S}_i =  \frac{|\dot{\bm{r}_i} - v_0 \bm{u}_i + D\nabla_i U|^2}{4D} + \frac{\dot\theta_i^2}{4D_r}
- \frac{D}{2} \nabla_i^2 U
\eeq
[We have neglected here a contribution to $P(X)$ from the initial condition, this does not cause a problem because we consider finally the long-time limit in (\ref{equ:I-bound-upper}).]

Defining the corresponding quantity ${\cal S}_i^{\rm con}$ for the controlled process one has
\beq
{\cal D}_{\rm KL}(P^{\rm con}||P)  = N\tau \langle {\cal S}_i - {\cal S}_i^{\rm con} \rangle_{\rm con}
\label{equ:DKL-S}
\eeq
where the average is computed in the steady state of the controlled process.  One finds
\begin{multline}
{\cal S}_i - {\cal S}_i^{\rm con}
= \frac{v^{\rm con} - v_0}{2D} \bm{u}_i \cdot \dot{\bm{r}}_i + \Delta^{\rm con}_i
\\ - \frac{|(v^{\rm con} - v_0) \bm{u}_i - D\nabla_i U^{\rm con} |^2}{4D}  + \frac{D}{2} \nabla^2 U^{\rm con}
\\ - \frac{1}{4D_r} {\left(  \frac{\partial U^{\rm con}}{\partial\theta_i} \right)^2} + \frac{D_r}{2} \left( \frac{\partial^2 U^{\rm con}}{\partial\theta_i^2} \right)
\label{equ:dS-con}
\end{multline}
where $\Delta^{\rm con}_i$ has the property that  $ \sum_i  \int_0^\tau \Delta^{\rm con}_i d\tau
= [ U^{\rm con}(0) - U^{\rm con}(\tau)]/2$ which will be negligible on taking the limit in (\ref{equ:I-bound-upper}).
The KL divergence is the average of this quantity with respect to the controlled dynamics -- the fact that $U^{\rm con}$ appears in both the action and in the equation of motion can be used to simplify the formulae for ${\cal D}_{\rm KL}$, as for example in Eq.~(H3) of~\cite{Nemoto2019}.  Such simplifications are not essential for this work, so we omit them.

For the controlled dynamics of (\ref{equ:u-con-g},\ref{equ:theta-con}), it is useful to denote by $\varphi$ the angle between $\bm{\nu}$ and the $x$-axis.  This can be obtained by considering
\beq
|\bm{\nu}| e^{\mathrm{i} \varphi} = \frac{1}{N} \sum_j e^{\mathrm{i} \theta_j}.
\label{equ:phi}
\eeq
Then the KL divergence of (\ref{equ:I-bound-upper}) can be obtained from (\ref{equ:DKL-S},\ref{equ:dS-con}), using $\nabla_i U^{\rm con}=0$ and $v^{\rm con}=v_0$, as
\beq
\lim_{\tau\to\infty} \frac{1}{N\tau} {\cal D}_{\rm KL}(P^{\rm con}_{g}||P)
=  \left\langle g \mathcal{I}_{1,\tau} - \frac{g^2}{D_r} \mathcal{I}_{2,\tau}  \right\rangle_{\rm con}  - \frac{g}{N}
\label{equ:DKL-con-g}
\eeq
where 
\begin{align}
\label{equ:torque-int1}
\mathcal{I}_{1,\tau} &= \frac{1}{\tau} \int_0^{\tau} |\bm{\nu}(t)|^2 \, \mathrm{d}t,\\
\label{equ:torque-int2}
\mathcal{I}_{2,\tau} &= \frac{1}{N \tau} \int_0^{\tau} |\bm{\nu}(t)|^2 \sum_{i} \sin(\theta_i(t) - \varphi(t))^2 \, \mathrm{d}t \; .
\end{align}
The steady-state averages of these integrals are independent of $\tau$; hence the right hand side of (\ref{equ:DKL-con-g}) is also independent of $\tau$ and no limit is required there.

The KL divergence of (\ref{equ:DKL-con-g}) was evaluated by numerical simulation of the controlled dynamics, to obtain the upper bounds in Fig.~\ref{fig:boundRate}.  (Analytic results are available for this quantity in the large-$N$ limit but numerical comparison with the large-deviation rate function requires results for finite $N$.)

\section{Cloning algorithm with modified dynamics}
\label{app:cloning}

\subsection{Modified dynamics}

The cloning algorithm is applied to ABPs similarly to~\cite{Nemoto2019}.  The number of clones is denoted by $n_c$.  We typically repeat each computation $N_{\rm runs}$ times, we take a simple average of the results and use the standard error for an error bar.
Convergence with respect to $n_c$ is discussed below.  The computational cost of our calculations is controlled primarily by $n_c$.

Given that we want to work with $n_c$ as small as possible, the low probabilities of large {deviation} events can be the origin of large systematic errors~\cite{Kurchan2006, Nemoto2016, Limmer2019}. To improve the convergence of the algorithm, we use a modification (or control) of the dynamics, informed by the typical behaviour of the system for large fluctuations of the biasing observable.
In particular we take (\ref{equ:dr-con-gen}) with $v^{\rm con} = v^{\rm con}_s$ from (\ref{equ:vcon-s}) and $U^{\rm con} = U_g$ from (\ref{equ:u-con-g}).  The choice of the parameter $g$ will be discussed below.

Following the same steps as (\ref{equ:dS-con}), we obtain the result of~\cite{Nemoto2019} that the path probability distributions for the controlled and original dynamics are related as
\beq
{P[X]}  \exp(- s N \tau (w_{\tau} ) ] \propto {P^{\rm con}[X]}  \exp(- s N \tau w_{\tau}^{\rm mod} )
\label{equ:wmod}
\eeq
where $X$ denotes a trajectory and the modified active work obeys
\beq
s w_{\tau}^{\rm mod} = s \left( 1 - \frac{s D}{v_0^2} + w_{f,\tau} \right) - g \left(\frac{1}{N} - \mathcal{I}_{1,\tau} + \frac{g}{D_r} \mathcal{I}_{2,\tau}\right)
\eeq
where $w_{f,\tau}$ is the force part of the active work from~(\ref{equ:wf}), and the integrals $\mathcal{I}_{1, \tau}$ and $\mathcal{I}_{2, \tau}$ are defined in (\ref{equ:torque-int1},\ref{equ:torque-int2}).

Physically, this means that the biased ensemble (\ref{equ:ave-bias}) for the ABP model (\ref{equ:dr}) can be formulated alternatively as a biased ensemble for the controlled ABP model, with  a modified bias $sw^{\rm mod}$.  Hence, the cloning algorithm is valid as a method for sampling large deviations for any choice of the parameter $g$ (including $g=0$).

\begin{figure}
\centering
\begin{overpic}[trim=0 0 0 0, width=0.48\textwidth]{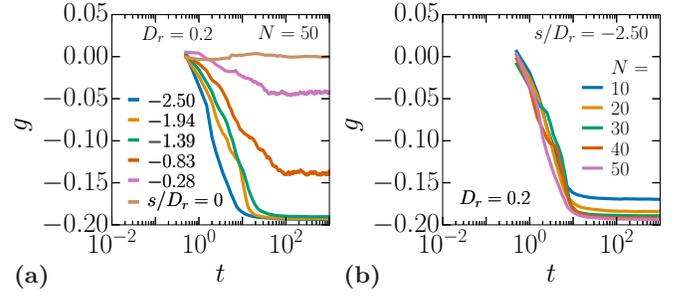}
\put (0, 0) {\texttt{\bf (a)}}
\put (\xp, 0) {\texttt{\bf (b)}}
\end{overpic}
\caption{Torque parameter $g$ as a function of time using the relation \eqref{equ:g-cloning} and setting $g(t=0) = 0$, for different {\bf (a)} biasing parameters $s$ and {\bf (b)} number of particles $N$. \textit{Parameter values}: $\phi=0.65$, $t_{\mathrm{max}} = 10^3$, $n_c=10^3$.}
\label{fig:conv-g}
\end{figure}

The role of the parameter $g$ is to improve the numerical efficiency, and hence to obtain accurate results with smaller numbers of clones.
Several methods have been proposed for determining suitable values of such parameters~\cite{Nemoto2016, Limmer2018, Limmer2019}.  Here we adopt the following method, similar to~\cite{Nemoto2019}, where the value is chosen ``on-the-fly'' within the algorithm.
We note that for the controlled dynamics with potential $U_g$, we have for large $N$ (and $g<D_r$) that 
\beq
\langle|\bm{\nu}|^2\rangle_{\rm con} = \frac{1}{N} \frac{1}{1 - g D_r^{-1}},
\label{equ:nu2-para}
\eeq
The derivation is given in Appendix~\ref{app:MFrotors}.
We obtained good results from the cloning algorithm by inverting this equation for $g$ and taking
\beq
g = D_r \left(1 - \frac{1}{N t^{-1} \langle\int_0^t \mathrm{d}t^{\prime} \, |\bm{\nu}(t^{\prime})|^2\rangle_{\rm clo}} \right)
\label{equ:g-cloning}
\eeq
where $\langle\ldots\rangle_{\rm clo}$ designates an average over the clones within the algorithm. Fig.~\ref{fig:conv-g} shows examples of such torque parameters as functions of the simulation time.
Note, this always gives $g<D_r$, the natural dynamics of the controlled system is always in the paramagnetic phase.  However, as the biased system moves into the ferromagnetic phase we find that $g$ gets close to $D_r$, and the controlled system approaches the mean-field critical point where fluctuations of $\bm{\nu}$ are large.  These large fluctuations are helpful for the  algorithm, in that they generate a wide range of trajectories, from which the cloning part of the algorithm can select those with large values of $w_\tau$.

\subsection{Convergence with respect to the number of clones}

\begin{figure}
\centering
\begin{overpic}[trim=0 0 0 0, width=0.48\textwidth]{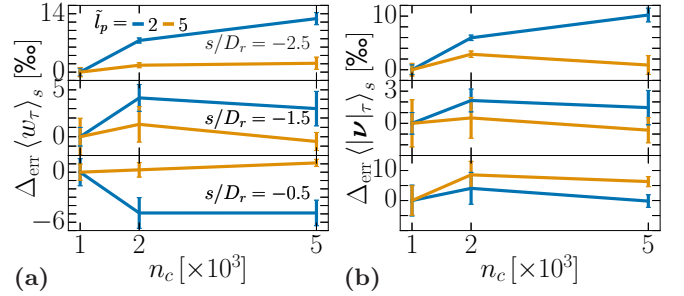}
\put (0, 0) {\texttt{\bf (a)}}
\put (\xp, 0) {\texttt{\bf (b)}}
\end{overpic}
\caption{Relative error of the value active work {\bf (left)} and order parameter {\bf (right)} to their value at $n_c = 10^3$ (leftmost value), $\Delta_{\mathrm{err}} \langle \bullet \rangle_s = \left[\langle \bullet \rangle_s - \langle \bullet \rangle_s(n_c=10^3)\right]/\langle \bullet \rangle_s(n_c=10^3)$. \textit{Parameter values}: $N=50$, $\phi=0.65$, $t_{\mathrm{max}} = 10^3$, $N_{\mathrm{runs}}=10$.}
\label{fig:clones}
\end{figure}

As stated in Ref.~\cite{Nemoto2016} the accuracy of the cloning algorithm is limited by the number $n_c$ of copies of our system which we simultaneously evolve. Most of the results presented in the main text where computed for $n_c = 10^3$ which is significantly lower than what was used for example in Ref.~\cite{Nemoto2019} (see Appendix~D of that work).

We show in Fig.~\ref{fig:clones} relative errors on the values of the active work and the order parameter, for two values of the persistence length and for $N=50$ particles, when varying the number of clones from $n_c=10^3$ to $n_c=5 \cdot 10^3$, with respect to their value for the former number of clones. We have that the behaviour of the relative error on both these quantities, which is at most of the order of 1\%, indicates that the qualitative conclusions of the main text are robust.

\section{Biased trajectories of two RTPs on a ring}
\label{app:rtp}

We consider the system of two RTPs defined in Sec.~\ref{sec:two-rtps}.
The SCGF of (\ref{equ:rtp-cgf}) is the largest eigenvalue $\psi^{\rm RTP}(\lambda)$ that solves
\beq
\psi^{\rm RTP}(\lambda) \bm{P}_{\lambda} = (\bm{\mathcal{L}} - \lambda \dot{w}^{\rm RTP}_f \bm{I}) \bm{P}_{\lambda}
\label{equ:rtp-eigenproblem}
\eeq
where $\bm{\mathcal{L}}$ is the Fokker-Planck operator acting on a probability distribution vector $\bm{P}_{\lambda} \equiv (P^{++}_{\lambda}, P^{--}_{\lambda}, P^{+-}_{\lambda}, P^{-+}_{\lambda})$.
The vector $\bm{P}_{\lambda}$ depends on the distance $r$ between the particles, which is measured clockwise around the circular system, starting at particle $1$.    Hence $0\leq r \leq L$. 
Since $\bm{P}$ is a vector, the operator $\bm{\mathcal{L}}$ is a $4\times 4$ matrix, and the Fokker-Planck equation is
\beq
\dot{P}^{\alpha_1\alpha_2} = \sum_{\alpha_1^{\prime},\alpha_2^{\prime} = \pm 1} \mathcal{L}^{\alpha_1\alpha_2,\alpha_1^{\prime}\alpha_2^{\prime}} P^{\alpha_1^{\prime}\alpha_2^{\prime}} \;
\eeq
 where the matrix elements of $\bm{\mathcal{L}}$ can be read off from the
 Fokker-Planck equation that corresponds to (\ref{equ:eom-rtp}), which is
\beq
\begin{aligned}
\dot{P}^{\alpha_1\alpha_2} =~& v_0(\alpha_1 - \alpha_2) \frac{\partial}{\partial r} P^{\alpha_1\alpha_2}
+ 2 \frac{\partial}{\partial r}\left(P^{\alpha_1\alpha_2}\frac{\partial}{\partial r} V\right)\\
& + \rtptime^{-1} \left(P^{\overline{\alpha}_1\alpha_2} + P^{\alpha_1\overline{\alpha}_2} - 2 P^{\alpha_1\alpha_2}\right)
\label{equ:rtp-fp}
\end{aligned}
\eeq
in which $\overline{\alpha}_i = - \alpha_i$. In the hard-core limit, we expect the probability density functions to take the form \cite{Slowman2016,Arnoulx2019}
\beq
\begin{aligned}
P^{\alpha_1\alpha_2}_{\lambda}(r) =~&\varepsilon^{\alpha_1\alpha_2}_{\lambda}(r)
+ \gamma^{\alpha_1\alpha_2,{\rm l}}_{\lambda} \delta(r) + \gamma^{\alpha_1\alpha_2,{\rm r}}_{\lambda} \delta(L - r) \; .
\end{aligned}
\label{equ:rtp-pdf-general}
\eeq
Here $\varepsilon^{\alpha_1\alpha_2}_{\lambda}$ is a smooth function of $r$ that describes the probability to find the particles with the given orientations, at a separation $r$. Also $\gamma^{\alpha_1\alpha_2,{\rm l}}_{\lambda}$ and $\gamma^{\alpha_1\alpha_2,{\rm r}}_{\lambda}$ indicate the probability that the particles are touching with particle 1 to the left (l) or right (r) of particle 2, with the prescribed orientations.

Particles never remain touching if their velocities point away from each other
so $\gamma^{+-,{\rm r}}_{\lambda} = \gamma^{-+,{\rm l}}_{\lambda} = 0$.
Note however that particles may be touching but moving parallel to each other, $\gamma^{++,{\rm r}}_{\lambda}>0$ in general.
In the hard core limit we have from (\ref{equ:rtp-bias}) that $\dot{w}^{\rm RTP}_f$ is non-zero only when particles are touching and have opposite orientation, so
\begin{align}
  \dot{w}^{\rm RTP}_f P^{+-}_{\lambda} & = -2v_0^2\gamma_\lambda^{+-,{\rm l}} \delta(r) \nonumber \\
  \dot{w}^{\rm RTP}_f P^{-+}_{\lambda} & =  -2v_0^2\gamma_\lambda^{-+,{\rm r}} \delta(L-r)
  \label{equ:w-contact}
\end{align}
with $\dot{w}^{\rm RTP}_f P^{++}_{\lambda} = 0 = \dot{w}^{\rm RTP}_f P^{--}_{\lambda}$.

The dominant eigenvector for (\ref{equ:rtp-eigenproblem}) obeys symmetry relations stemming from particle interchangeability (\ref{equ:rtp-sym-interchangeability}) and parity symmetry (\ref{equ:rtp-sym-parity}):
\begin{align}
\label{equ:rtp-sym-interchangeability}
P^{\alpha_1\alpha_2}_{\lambda}(r) &= P^{\alpha_2\alpha_1}_{\lambda}(L - r)\\
\label{equ:rtp-sym-parity}
P^{\alpha_1\alpha_2}_{\lambda}(r) &= P^{\overline{\alpha}_1\overline{\alpha}_2}_{\lambda}(L - r) \; .
\end{align}
We take as normalisation condition
\beq
\int_0^L \sum_{\alpha_1, \alpha_2 = \pm 1} P^{\alpha_1\alpha_2}_{\lambda}(r) \, \mathrm{d}r = 1 \; .
\label{equ:rtp-normalisation}
\eeq

The symmetry relations (\ref{equ:rtp-sym-interchangeability}, \ref{equ:rtp-sym-parity}) imply for the probability density function of aligned particles that
\begin{align}
\varepsilon^{++}_{\lambda}(r) = \varepsilon^{--}_{\lambda}(r) &= \varepsilon^{\alpha\alpha}_{\lambda}(r)\\
\label{equ:rtp-alpha-alpha-parity}
\varepsilon^{\alpha\alpha}_{\lambda}(r) &=  \varepsilon^{\alpha\alpha}_{\lambda}(L - r) \; .
\end{align}
Similarly for the ``sticking'' terms
\begin{align}
\gamma^{+-,{\rm r}}_{\lambda} = \gamma^{-+,{\rm l}}_{\lambda} &= \gamma^{\alpha\overline{\alpha}}_{\lambda}\\
\gamma^{\alpha\alpha,{\rm l}}_{\lambda} = \gamma^{\alpha\alpha,{\rm r}}_{\lambda} &= \gamma^{\alpha\alpha}_{\lambda}
\label{equ:rtp-stick-def}
\end{align}
which are to be interpreted as definitions of $\gamma^{\alpha\overline{\alpha}}_{\lambda},\gamma^{\alpha\alpha}_{\lambda}$.
These symmetry relations greatly simplify our calculations.

We highlight that the introduction of thermal diffusion in Eq.~\ref{equ:eom-rtp} would smooth out the $\delta$-functions associated to sticking in Eq.~\ref{equ:rtp-pdf-general}, which will be replaced by exponentials decaying away from contact on a length scale set by the diffusivity \cite{Das2020}.

\subsection{Unbiased steady state distribution}

We first solve the case without bias ($\lambda = 0$) in steady state ($\dot{\bm{P}} = 0$).
Evaluating Eq.~\ref{equ:rtp-fp} for $r\neq 0,L$,  gives
\beq
0 = 2 v_0 \frac{\partial}{\partial r} \varepsilon^{\alpha\overline{\alpha}}_0(r) + \rtptime^{-1} [2 \varepsilon^{\alpha\alpha}_0(r) - 2 \varepsilon^{\alpha\overline{\alpha}}_0(r)] \; .
\eeq
Using
the symmetry properties~(\ref{equ:rtp-sym-interchangeability},~\ref{equ:rtp-sym-parity}) leads to $\partial_r \varepsilon^{\alpha\overline{\alpha}}_0(r) = 0$ and hence $\varepsilon^{\alpha\overline{\alpha}}_0(r) = \varepsilon^{\alpha\alpha}_0(r) = \varepsilon_0$.  In other words, $P$ is independent of $r$ and of the orientations, except when the particles are touching.

Relations between $\gamma^{\alpha\alpha}_0$ and $\gamma^{\alpha\overline{\alpha}}_0$ follow from integrating Eq.~\ref{equ:rtp-fp} from $r=0^-$ to $r=\epsilon$ and then taking the hard-core limit:
\begin{align}
\rtptime^{-1}(\gamma^{\alpha\overline{\alpha}}_0 - 2 \gamma^{\alpha\alpha}_0) &= 0\\
- 2 v_0 \varepsilon_0 + \rtptime^{-1} 2 \gamma^{\alpha\alpha}_0 &= 0 \; .
\end{align}
Finally using the normalisation condition (\ref{equ:rtp-normalisation})
\beq
4 L \varepsilon_0 + 4 \gamma^{\alpha\alpha}_0 + 2 \gamma^{\alpha\overline{\alpha}}_0 = 1 \; .
\eeq
One can then solve to obtain
\begin{align}
\label{equ:rtp-P0-++}
P^{++}_0(r) = P^{--}_0(r) & = a + la \delta(r) + la \delta(L -r) \\
\label{equ:rtp-P0-+-}
P^{+-}_0(r) & = a + 2 la \delta(r)
\\
 P^{-+}_0(r) & = a + 2 la \delta(L-r)
\end{align}
where $a = [4(L + 2l)]^{-1}$, and the persistence length $l = v_0\rtptime$, as in the main text.  This exactly recovers the continuous space and time results of Ref.~\cite{Slowman2016} (Eqs.~11,~12).  Also
\beq
\left\langle \dot{w}^{\rm RTP}_f \right\rangle = - \frac{2 l v_0^2}{L + 2l}
\label{equ:rtp-mean-w}
\eeq
by (\ref{equ:w-contact}).

Note that a steady state distribution that is non-uniform also for particles not in contact, as observed for RTPs on a discrete lattice \cite{Slowman2016}, can be introduced by considering finite-time tumbles \cite{Slowman2017} or thermal diffusion \cite{Das2020}.

\subsection{Biased steady state distribution}

We now solve the general biased case ($\lambda \in \mathbb{R}$) using (\ref{equ:rtp-eigenproblem}).  The method follows the unbiased case.
For particles not in contact we have
\begin{align}
\label{equ:rtp-eigenproblem-++}
&\begin{aligned}
&\psi^{\rm RTP}(\lambda) \varepsilon^{\alpha\alpha}_{\lambda}(r) =\\
&\qquad\rtptime^{-1} (\varepsilon^{\alpha\overline{\alpha}}_{\lambda}(r) + \varepsilon^{\alpha\overline{\alpha}}_{\lambda}(L - r) - 2 \varepsilon^{\alpha\alpha}_{\lambda}(r))
\end{aligned}\\
\label{equ:rtp-eigenproblem-+-}
&\begin{aligned}
&\psi^{\rm RTP}(\lambda) \varepsilon^{\alpha\overline{\alpha}}_{\lambda}(r) =  2 \alpha v_0 \frac{\partial}{\partial r} \varepsilon^{\alpha\overline{\alpha}}_{\lambda}(r)\\
&\qquad+ \rtptime^{-1} (2 \varepsilon^{\alpha\alpha}_{\lambda}(r) - 2 \varepsilon^{\alpha\overline{\alpha}}_{\lambda}(r))
\end{aligned}
\end{align}
Hence
\beq
 \frac{{\rm d}^2}{{\rm d}r^2} ( \varepsilon_\lambda^{\alpha\overline{\alpha}} + \varepsilon_\lambda^{\overline{\alpha}\alpha} )
 =  k_\lambda^2( \varepsilon_\lambda^{\alpha\overline{\alpha}} + \varepsilon_\lambda^{\overline{\alpha}\alpha} )
\eeq
with 
\beq
k_{\lambda}^2 l^2 = \frac{ \rtptime \psi^{\rm RTP}(\lambda) }{4}  \left[ 4 + \rtptime \psi^{\rm RTP}(\lambda) \right] \; .
\label{equ:rtp-k}
\eeq
Note that $k_\lambda$ may be either real or imaginary.
The solutions for $\varepsilon$ (in both cases) can then be expressed as
\begin{align}
\label{equ:rtp-epsilon-++}
&
\varepsilon^{\alpha\alpha}_{\lambda}(r) 
= \frac{1}{2 + \rtptime \psi^{\rm RTP}(\lambda)}
A_{\lambda} (e^{-k_{\lambda} r} + e^{-k_{\lambda} (L - r)})
\\
\label{equ:rtp-epsilon-+-}
&\begin{aligned}
2\varepsilon^{\alpha\overline{\alpha}}_{\lambda}(r) =& \left(1 - \frac{2\alpha  k_{\lambda} l}{2 + \rtptime\psi^{\rm RTP}(\lambda)}\right) A_{\lambda} e^{- k_{\lambda} r}\\
&+\left(1 + \frac{2\alpha  k_{\lambda} l}{2 + \rtptime\psi^{\rm RTP}(\lambda)}\right) A_{\lambda} e^{- k_{\lambda} (L - r)}
\end{aligned}
\end{align}
where there is a single constant of integration $A_{\lambda}$ because we have enforced the symmetry (\ref{equ:rtp-alpha-alpha-parity}).

We now derive four equations that can be used to express $(\gamma^{\alpha\alpha}_{\lambda}, \gamma^{\alpha\overline{\alpha}}_{\lambda}, A_{\lambda}, \lambda)$ as functions of $\psi^{\rm RTP}(\lambda)$, which enables a full solution of this problem.

Integrating Eq.~\ref{equ:rtp-eigenproblem} from $0^-$ to $\epsilon$, as in the unbiased case,
and taking the $++$ component of the vector $\bm{P}$, one obtains
\beq
\label{equ:rtp-gamma-relation}
\psi^{\rm RTP}(\lambda) \gamma^{\alpha\alpha}_{\lambda} = \rtptime^{-1} (\gamma^{\alpha\overline{\alpha}}_{\lambda} - 2 \gamma^{\alpha\alpha}_{\lambda}) \; .
\eeq
Similarly, taking the $P^{-+}$ component gives
\beq
\begin{aligned}
0 =& - 2 v_0 \varepsilon^{-+}_{\lambda}(0^+) + \rtptime^{-1} 2 \gamma^{\alpha\alpha}_{\lambda} \; ,
\end{aligned}
\eeq
in which  $\varepsilon^{-+}_\lambda(0^+)$ may be substituted
using (\ref{equ:rtp-epsilon-+-}) to obtain
\beq
  \gamma^{\alpha\alpha}_{\lambda}
= \frac{\rtptime v_0 A_{\lambda}}{2} \Big[(1 + e^{-k_{\lambda} L})\\
+ \frac{2 k_{\lambda} l}{2 + \rtptime\psi^{\rm RTP}(\lambda)} (1 - e^{-k_{\lambda} L})\Big] \; .
\label{equ:rtp-A-gamma-boundary}
\eeq
In addition, using (\ref{equ:rtp-epsilon-++},\ref{equ:rtp-epsilon-+-})
in the normalisation condition (\ref{equ:rtp-normalisation}) leads to
\begin{multline}
1 = 2 \gamma^{\alpha\alpha}_{\lambda} \left[4 + \rtptime \psi^{\rm RTP}(\lambda) \right]\\
 + \frac{2A_{\lambda}}{k_{\lambda}} \frac{(1 - e^{-k_{\lambda}L})[4 + \rtptime\psi^{\rm RTP}(\lambda) ]}{2 + \rtptime\psi^{\rm RTP}(\lambda) }
\label{equ:rtp-normalisation-exact}
\end{multline}
where we have also used Eq.~\ref{equ:rtp-gamma-relation} to eliminate $\gamma^{\alpha\overline{\alpha}}$.

Now, since the Fokker-Planck equation (\ref{equ:rtp-fp}) preserves the normalisation of $\bm{P}$, one may integrate (\ref{equ:rtp-eigenproblem}) over $r$ and sum over $\alpha_1,\alpha_2$, then apply (\ref{equ:rtp-normalisation}) to obtain
$ \psi^{\rm RTP}(\lambda) = 4 \lambda v_0^2 \gamma^{\alpha\overline{\alpha}}_{\lambda}$.  Then use (\ref{equ:rtp-gamma-relation}) to obtain
\beq
\begin{aligned}
\psi^{\rm RTP}(\lambda)
&= 4 \gamma^{\alpha\alpha}_{\lambda} \lambda v_0^2 \left[ 2+ \rtptime \psi^{\rm RTP}(\lambda)  \right]  \; .
\end{aligned}
\label{equ:rtp-psi-gamma}
\eeq

Eqs.~(\ref{equ:rtp-gamma-relation},\ref{equ:rtp-A-gamma-boundary},\ref{equ:rtp-normalisation-exact},\ref{equ:rtp-psi-gamma}) are the promised four equations for
$(\gamma^{\alpha\alpha}_{\lambda}, \gamma^{\alpha\overline{\alpha}}_{\lambda}, A_{\lambda}, \lambda)$, in terms of $\psi^{\rm RTP}(\lambda)$.
The problem is now solved by computing the inverse of $\psi^{\rm RTP}(\lambda)$, which amounts to treating $\psi^{\rm RTP}$ as a parameter and solving for $\lambda$ and the other variables.  Note that $k_\lambda$ is fully determined by the value of $\psi^{\rm RTP}$, see (\ref{equ:rtp-k}).

To simplify the computation, we
introduce dimensionless variables that treat the persistence length $l$ as the unit of length: 
\begin{align}
\tilde{\lambda} &= \lambda l v_0\nonumber\\
\tilde{A}_{\tilde\lambda} &= A_{\lambda} l\nonumber\\
\tilde{\psi}^{\rm RTP} &= \rtptime \psi^{\rm RTP}\\
\tilde{k}_{\lambda} &= k_{\lambda} l \nonumber \\
\tilde{L} &= \frac{L}{l}\nonumber
\end{align}
The ratio $\gamma^{\alpha\alpha}_{\tilde{\lambda}}/\tilde{A}_{\tilde{\lambda}}$ determines the relative probabilities of the particles being in contact or separated.  
One has from (\ref{equ:rtp-A-gamma-boundary}) that this ratio can be expressed in terms of $\tilde\psi^{\rm RTP}$ as
\beq
\label{equ:rtp-A-gamma-exact}
\frac{\gamma^{\alpha\alpha}_{\tilde{\lambda}}}
{\tilde{A}_{\tilde{\lambda}}}
 = \frac{1 + e^{-\tilde{k}_{\tilde{\lambda}} \tilde{L}}}{2}
 + \frac{ \tilde{k}_{\tilde{\lambda}} (1 - e^{-\tilde{k}_{\tilde{\lambda}} \tilde{L}})}
        {2 + \tilde{\psi}^{\rm RTP}}
\eeq
and, dividing Eq.~\ref{equ:rtp-normalisation-exact} by Eq.~\ref{equ:rtp-psi-gamma} yields (after some rearrangements):
\beq
\label{equ:rtp-lambda-A-gamma-exact}
\tilde{\lambda} = 
\frac{\tilde{\psi}^{\rm RTP}(\tilde{\psi}^{\rm RTP} + 4)}{2(\tilde{\psi}^{\rm RTP} + 2)}
\bigg[ 1 +
\frac{\tilde{A}_{\tilde{\lambda}}}{\gamma^{\alpha\alpha}_{\tilde{\lambda}}}
\frac{(1 - e^{-\tilde{k}_{\lambda} \tilde{L}})}
{\tilde{k}_{\tilde{\lambda}} (\tilde{\psi}^{\rm RTP} + 2)}
\bigg]
\eeq
yielding $\tilde{\psi}^{\rm RTP}(\tilde{\lambda}) \sim 2 \tilde{\lambda}$ when $\lambda \to \infty$.
Combining (\ref{equ:rtp-A-gamma-exact},~\ref{equ:rtp-lambda-A-gamma-exact}) gives the promised inverse of $\tilde\psi^{\rm RTP}(\tilde\lambda)$, as
\beq
\label{equ:rtp-lambda-psi-exact}
\tilde{\lambda} =
\frac
{\tilde{\psi}^{\rm RTP}(\tilde{\psi}^{\rm RTP} + 4)}
{(\tilde{\psi}^{\rm RTP} + 2)} \left[ \frac12 + \frac{1}{ \Omega_{\tilde L}(\tilde{\psi}^{\rm RTP}) }
\right]
\eeq
with
\begin{multline}
\Omega_{\tilde L}(\tilde{\psi}^{\rm RTP})
 =
\frac{\tilde{\psi}^{\rm RTP}}{2} (\tilde{\psi}^{\rm RTP} + 4)
\\ + \tilde{k}_{\tilde{\lambda}} (\tilde{\psi}^{\rm RTP} + 2) \frac{1 + e^{-\tilde{k}_{\tilde{\lambda}} \tilde{L}}}{1 - e^{-\tilde{k}_{\tilde{\lambda}} \tilde{L}}} \; ,
\end{multline}
where we used also (\ref{equ:rtp-k}).

Recall from (\ref{equ:rtp-k}) that
$\tilde{k}$ is an imaginary number
for $-4 < \tilde{\psi}^{\rm RTP} < 0$.
In this case it is useful to rewrite
\beq
\tilde{k}_{\tilde{\lambda}} \frac{1 + e^{-\tilde{k}_{\tilde{\lambda}} \tilde{L}}}{1 - e^{-\tilde{k}_{\tilde{\lambda}} \tilde{L}}}
=
|\tilde{k}_{\tilde{\lambda}}| \cot\left(\frac{|\tilde{k}_{\tilde\lambda}| \tilde{L}}{2}\right)
\eeq
With this result in hand, a careful analysis shows that
$\Omega_{\tilde L}(\psi)$ has at least one zero for $-2 < \psi < 0$, at which point $\tilde\lambda$ diverges.  This implies that $\tilde{\psi}^{\rm RTP}(\tilde{\lambda})$ has a horizontal tangent for $\tilde{\lambda} \to -\infty$.  The location of the (largest) zero sets the smallest possible value for $\tilde{\psi}^{\rm RTP}(\tilde{\lambda})$, which is achieved as $\tilde\lambda\to-\infty$.  See Fig.~\ref{fig:rtp}(a).

\subsection{Scaling regime}
\label{app:rtp-scaling}

\begin{figure}
\centering
\includegraphics[width=0.48\textwidth]{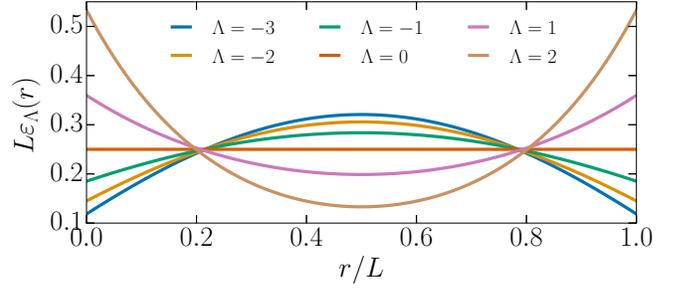}
\caption{Regular part of the probability density function $\varepsilon_{\Lambda}(r)$ from (\ref{equ:rtp-epsilon-pos},\ref{equ:rtp-epsilon-neg}) scaled by the ring length $L$.}
\label{fig:rtp-scaling}
\end{figure}

It is instructive to consider two particles in a very large system, ${L}\to\infty$.
The system has an associated scaling limit whose behaviour is shown in Fig~\ref{fig:rtp-scaling}. 

Since the persistence length of the run-and-tumble motion is much less than the system size, the particle motion on large scales can be characterised as (athermal) diffusion, and the particle explores the system on a time scale $O(L^2)$.  Since
 $\psi^{\rm RTP}$ is an inverse time scale, it is expected to be $O(L^{-2})$.
Moreover, it follows from Eq.~(\ref{equ:rtp-mean-w}) that typical values of $w_f^{\rm RTP}$ are of order $L^{-1}$ in this regime.  Physically, this small value arises because
of the small fraction of time that the particles spend in contact, when $L$ is large.  Hence
\beq
\psi^{\rm RTP}(\lambda) = \frac{2 l v_0^2}{L + 2l} \lambda + \mathcal{O}(\lambda^2) \; .
\eeq
One then expects a scaling form as $L\to\infty$:
\beq
\psi^{\rm RTP}(\lambda) \simeq L^{-2} \varphi(\lambda L) \; ,
\eeq
which will be verified below.  The corresponding form of the rate function for $w_f^{\rm RTP}$ is obtained from (\ref{equ:I-legendre}) as
\beq
I(w_f) \simeq L^{-2} {\cal I}(w_fL) \; .
\eeq

The natural dimensionless quantities in this regime are
\begin{align}
\label{equ:rtp-Lambda}
\Lambda &= \frac{\lambda L v_0}{2}\\
\label{equ:rtp-Phi}
\Psi(\Lambda) &= \frac{ \rtptime L^2}{4 l^2}  \psi^{\rm RTP}(\lambda)
\end{align}
so that $\varphi(\lambda L) = 2 l^2\Psi(\Lambda)/\rtptime$.  The quantities $\Lambda,\Psi(\Lambda)$ have same sign.

Since $\psi^{\rm RTP}$ is $O(L^{-2})$ then Eq.~\ref{equ:rtp-lambda-psi-exact} shows that $k=O(L^{-1})$.
At the lowest order in $L^{-1}$ we then infer from Eq.~\ref{equ:rtp-lambda-psi-exact}
that for $\Lambda>0$ then
\beq 
\label{equ:rtp-Lambda-pos}
{\Lambda} 
= \sqrt{\Psi}\tanh\left(\sqrt{\Psi}\right),
\eeq
while for $\Lambda<0$ we have
\beq
\label{equ:rtp-Lambda-neg}
{\Lambda} 
= -\sqrt{|\Psi|}\tan\left(\sqrt{|\Psi|}\right) 
\eeq 
yielding $\lim_{\Lambda \to -\infty} \Psi(\Lambda) = -\pi^2/4$.

Moreover, at leading order in $L^{-1}$, it follows from Eqs.~\ref{equ:rtp-epsilon-++},~\ref{equ:rtp-epsilon-+-} that
\beq
\begin{aligned}
&\varepsilon^{\alpha\alpha}_{\Lambda}(r) = \varepsilon^{\alpha\overline{\alpha}}_{\Lambda}(r) = \varepsilon_{\Lambda}(r)\\
&\qquad = \frac{1}{2} A_{\Lambda} \left(e^{-k_{\Lambda} r} + e^{-k_{\Lambda} (L - r)}\right)
\end{aligned}
\eeq
and from Eqs.~\ref{equ:rtp-psi-gamma},~\ref{equ:rtp-A-gamma-exact} that
\begin{align}
\label{equ:rtp-sticking}
\gamma_{\Lambda} &= \frac{l}{4 L} \frac{\Psi(\Lambda)}{\Lambda}\\
A_{\Lambda} &= \frac{2 \gamma_{\Lambda}}{l} \frac{1}{1 + e^{-k_{\Lambda} L}} \; .
\end{align}
Using Eq.~\ref{equ:rtp-k} leads for $\Lambda>0$ to
\beq
\varepsilon_{\Lambda}(r) = \frac{1}{4 L} \frac{\Psi(\Lambda)}{\Lambda} \frac{\cosh\left(\sqrt{\Psi(\Lambda)}\left(1 - \frac{2r}{L}\right)\right)}{\cosh\left(\sqrt{\Psi(\Lambda)}\right)}
\label{equ:rtp-epsilon-pos}
\eeq
and for $\Lambda<0$ to
\beq
\varepsilon_{\Lambda}(r) = \frac{1}{4 L} \frac{\Psi(\Lambda)}{\Lambda}
\frac{\cos\left(\sqrt{|\Psi(\Lambda)|}\left(1 - \frac{2r}{L}\right)\right)}{\cos\left(\sqrt{|\Psi(\Lambda)|}\right)}
\label{equ:rtp-epsilon-neg}
\eeq
which we plot in Fig.~\ref{fig:rtp-scaling}.

The physical picture emerging from Fig.~\ref{fig:rtp-scaling} is as follows.  In a large system, a very weak bias $\lambda = O(L^{-1})$ is sufficient to change qualitatively the separation of the particles.  The resulting probability distributions are independent of particle orientation but depend on the particle separation through the scaling variable $r/L$.  For $\Lambda>0$ (corresponding to reduced active work), the particles are more likely to approach each other, which favours collisions.  For $\Lambda<0$ (enhanced work), they are more likely to be far apart, suppressing collisions.  One sees from (\ref{equ:rtp-sticking}) that the probability to find the particles in contact vanishes as $L^{-1}$, this holds throughout the scaling regime $\lambda=O(L^{-1})$.

\subsection{Distribution over the infinite-time interval}

The probability density vector $\bm{P}_{\lambda}$ which satisfies (\ref{equ:rtp-eigenproblem}, \ref{equ:rtp-normalisation}) indicates the fraction of trajectories for which the particles have final orientations $\alpha_i$ and separation $r$ in the $\lambda$-ensemble \cite{Nemoto2016, Chetrite2015}.
We are also interested in the fraction of time spent with given orientations $\alpha_i$ and separation $r$ in the $\lambda$-ensemble, which we will denote $\hat{\bm{P}}_{\lambda}$. We expect these probability density functions to take the same form and respect the same symmetries as their final time counterparts (\ref{equ:rtp-pdf-general}, \ref{equ:rtp-sym-interchangeability}, \ref{equ:rtp-sym-parity}), so that
\begin{align}
\hat{P}^{\alpha\alpha}_{\lambda}(r) & = \hat{\varepsilon}^{\alpha\alpha}_{\lambda}(r) + \hat{\gamma}^{\alpha\alpha}_{\lambda} \delta(r) + \hat{\gamma}^{\alpha\alpha}_{\lambda} \delta(L - r)\\
\hat{P}^{+-}_{\lambda}(r) & = \hat{\varepsilon}_{+-}(r) + \hat{\gamma}_{\alpha\overline{\alpha}} \delta(r)\\
\hat{P}^{-+}_{\lambda}(r) & = \hat{\varepsilon}_{-+}(r) + \hat{\gamma}_{\alpha\overline{\alpha}} \delta(L - r)
\end{align}
in addition to being normalised accordingly \eqref{equ:rtp-normalisation}.
In order to compute them it is necessary to solve the eigenproblem adjoint to \eqref{equ:rtp-eigenproblem}
\beq
\psi^{\rm RTP}(\lambda) \bm{Q}_{\lambda} = (\bm{\mathcal{L}}^{\dagger} - \lambda \dot{w}^{\rm RTP}_f \bm{I}) \bm{Q}_{\lambda}
\label{equ:rtp-adjoint-eigenproblem}
\eeq
where the same $\psi^{\rm RTP}(\lambda)$ is the largest eigenvalue. It then follows that
\beq
\hat{P}^{\alpha_1\alpha_2}_{\lambda}(r) = P^{\alpha_1\alpha_2}_{\lambda}(r) Q^{\alpha_1\alpha_2}_{\lambda}(r)
\eeq
according to Ref.~\cite{Chetrite2015}.

The matrix elements of $\bm{\mathcal{L}}^{\dagger}$ can be read off from the backward Fokker-Planck equation that corresponds to \eqref{equ:rtp-fp}
\beq
\begin{aligned}
\dot{Q}^{\alpha_1\alpha_2} = & -v_0(\alpha_1 - \alpha_2) \frac{\partial}{\partial r} Q^{\alpha_1\alpha_2}
- 2 \frac{\partial}{\partial r} Q^{\alpha_1\alpha_2} \frac{\partial}{\partial r} V\\
& + \rtptime^{-1} \left(Q^{\overline{\alpha}_1\alpha_2} + Q^{\alpha_1\overline{\alpha}_2} - 2 Q^{\alpha_1\alpha_2}\right)
\label{equ:rtp-bfp}
\end{aligned}
\eeq
and indicate that the eigenvector of this operator corresponding to the eigenvalue 0, which is also $\bm{Q}_0$ in \eqref{equ:rtp-adjoint-eigenproblem}, is constant.
Since bias introduces in \eqref{equ:rtp-adjoint-eigenproblem} terms which are at least as regular as those in \eqref{equ:rtp-bfp}, we expect $Q^{\alpha_1\alpha_2}_{\lambda}$ to remain smooth and continuous for $\lambda \neq 0$.
For particles not in contact, $Q^{\alpha_1\alpha_2}_{\lambda}$ satisfies the same equations as $P^{\alpha_2\alpha_1}_{\lambda}$ (\ref{equ:rtp-eigenproblem-++}, \ref{equ:rtp-eigenproblem-+-}), with the replacement $v_0\to -v_0$. We can then finally conclude
\begin{align}
\label{equ:rtp-ave-eaa}
\hat{\varepsilon}^{\alpha\alpha}_{\lambda}(r) & = \frac{\hat{A}_{\lambda}}{A_{\lambda}} \varepsilon^{\alpha\alpha}_{\lambda}(r)^2\\
\label{equ:rtp-ave-eaba}
\hat{\varepsilon}^{\alpha\overline{\alpha}}_{\lambda}(r) & = \frac{\hat{A}_{\lambda}}{A_{\lambda}} \varepsilon^{\alpha\overline{\alpha}}_{\lambda}(r) \varepsilon^{\overline{\alpha}\alpha}_{\lambda}(r)\\
\label{equ:rtp-ave-gaa}
\hat{\gamma}^{\alpha\alpha}_{\lambda} & =\frac{\hat{A}_{\lambda}}{A_{\lambda}}  \gamma^{\alpha\alpha}_{\lambda}  \varepsilon^{\alpha\alpha}_{\lambda}(0^+)\\
\label{equ:rtp-ave-gaba}
\hat{\gamma}^{\alpha\overline{\alpha}}_{\lambda} & = \frac{\hat{A}_{\lambda}}{A_{\lambda}} \gamma^{\alpha\overline{\alpha}}_{\lambda}  \varepsilon^{+-}_{\lambda}(L^-)
\end{align}
where $\hat{A}_{\lambda}$ is a normalisation constant for $\hat{P}_{\lambda}$.

\subsection{Polarisation}

From (\ref{equ:nu-rtp-ave-nubar},\ref{equ:nu-rtp-ave-nuend}) we identify
\begin{align}
\label{equ:nu-rtp-ave}
\nu^{\rm RTP}_{\rm ave}(\lambda) = \int_0^L \left(\hat{P}^{++}_{\lambda}(r) + \hat{P}^{--}_{\lambda}(r)\right) \, \mathrm{d}r
\end{align}
and (replacing $\hat{P}$ with ${P}$)
\beq
\label{equ:nu-rtp-end}
\nu^{\rm RTP}_{\rm end}(\lambda)
= \int_0^L \left(P^{++}_{\lambda}(r) + P^{--}_{\lambda}(r)\right) \, \mathrm{d}r \; .
\eeq

An exact expression of the polarisation $\nu^{\rm RTP}_{\rm end}$ (\ref{equ:nu-rtp-end}) can be computed by noting that
\beq
\nu^{\rm RTP}_{\rm end}(\lambda) = \frac{4\gamma^{\alpha\alpha}_{\lambda} +
2\int_0^L \varepsilon^{\alpha\alpha}_{\lambda}(r) \, \mathrm{d}r }
{ 4\gamma^{\alpha\alpha}_{\lambda} + 2\gamma^{\alpha\overline{\alpha}}_{\lambda} +
2\int_0^L [\varepsilon^{\alpha\alpha}_{\lambda}(r) + \varepsilon^{\alpha\overline{\alpha}}_{\lambda}(r) ] \, \mathrm{d}r }
\eeq
where the denominator is $1$ by (\ref{equ:rtp-normalisation}) and we used that $\int_0^L [ \varepsilon^{\alpha\overline{\alpha}}_{\lambda}(r) - \varepsilon_\lambda^{\overline{\alpha}\alpha}(r) ] dr =0$. Moreover, Eq.~\ref{equ:rtp-gamma-relation} yields $\gamma^{\alpha\overline{\alpha}}_{\lambda} = (\tilde{\psi}^{\rm RTP} + 2) \gamma^{\alpha\alpha}_{\lambda}$ and Eqs.~\ref{equ:rtp-epsilon-++}, \ref{equ:rtp-epsilon-+-} yield
\beq
2 \int_0^L \varepsilon^{\alpha\overline{\alpha}}_{\lambda}(r) \, \mathrm{d}r = (\tilde{\psi}^{\rm RTP} + 2) \int_0^L \varepsilon^{\alpha\alpha}_{\lambda}(r) \, \mathrm{d}r
\eeq
therefore
\beq
\nu^{\rm RTP}_{\rm end} = \frac{2}{ \tilde{\psi}^{\rm RTP} + 4 }
\label{equ:nu-rtp-end-exact}
\eeq
whose dependence on $\tilde{\lambda}$ can then be obtained parametrically via (\ref{equ:rtp-lambda-psi-exact}).

An exact expression of the polarisation $\nu^{\rm RTP}_{\rm ave}$ (\ref{equ:nu-rtp-ave}) is also available by writing
\begin{widetext}
\beq
\nu^{\rm RTP}_{\rm ave}(\lambda) = \frac{4 \gamma^{\alpha\alpha}_{\lambda} \varepsilon^{\alpha\alpha}_{\lambda}(0^+) + 2 \int_0^L \varepsilon^{\alpha\alpha}_{\lambda}(r)^2 \, \mathrm{d}r}{4 \gamma^{\alpha\alpha}_{\lambda} \varepsilon^{\alpha\alpha}_{\lambda}(0^+) + 2 \gamma^{\alpha\alpha}_{\lambda} (\tilde{\psi}^{\rm RTP} + 2) \varepsilon^{+-}_{\lambda}(L^-) + 2 \int_0^L \varepsilon^{\alpha\alpha}_{\lambda}(r)^2 \, \mathrm{d}r + 2 \int_0^L \varepsilon^{\alpha\overline{\alpha}}_{\lambda}(r)\varepsilon^{\overline{\alpha}\alpha}_{\lambda}(r) \, \mathrm{d}r}
\eeq
which avoids the need to determine $\hat{A}_{\lambda}$, and where we have used (\ref{equ:rtp-ave-eaa}, \ref{equ:rtp-ave-eaba}, \ref{equ:rtp-ave-gaa}, \ref{equ:rtp-ave-gaba}).
This allows $\nu^{\rm RTP}_{\rm ave}$ to be determined in full via (\ref{equ:rtp-epsilon-++}, \ref{equ:rtp-epsilon-+-}, \ref{equ:rtp-A-gamma-exact}).
\end{widetext}

It is easily verified that $\nu^{\rm RTP}_{\rm end} = \nu^{\rm RTP}_{\rm ave} = 1/2$ for $\lambda = 0$, corresponding to a state where aligned and anti-aligned states are equiprobable.

\section{Statistics of orientational order parameter(s)}
\label{app:largedev-JJ}

This Appendix analyses (controlled) ABPs in situations where their orientations evolve independently of their positions.  In these cases the statistics of the order parameter $\bm{\nu}$ can be computed.

\subsection{Mean-field analysis}
\label{app:MFrotors}

We consider the dynamics of the particle orientations alone, for the controlled system
(\ref{equ:u-con-g}).
We have $U^{\rm con}_g=-gN^{-2} \sum_{ij} \cos(\theta_i-\theta_j)$ so the controlled equation of motion for the orientation is
\beq
\dot\theta_i = gN^{-2} \sum_j \sin(\theta_i-\theta_j) + \sqrt{2 D_r} \xi_i
\eeq
Writing $ \sin(\theta_i-\theta_j) = \sin\theta_i \cos\theta_j - \cos\theta_i \sin\theta_j$, and using (\ref{equ:def-nu}) with $\bm{\nu} = |\bm{\nu}| ( \cos\varphi,\sin\varphi)$ yields (\ref{equ:theta-con}) of the main text.

For $N \gg 1$, mean-field theory is valid, because the individual $\theta_i$ relax much faster than the global $\varphi$, and  fluctuations of $\bm{\nu}$ are also negligible.
Without loss of generality, we take $\varphi = 0$, hence it is consistent to set
\beq
|\bm{\nu}| = \langle \cos\theta_i \rangle_{\rm con}
\label{equ:MFnu}
\eeq
in (\ref{equ:theta-con}).
Treating this quantity as a fixed number,
the steady state of the system obeys a Boltzmann distribution where each $\theta_i$ is independent with distribution
\beq
p^{\rm con}_{|\bm{\nu}| }(\theta_i ) \propto  \exp\left(\frac{2 g |\bm{\nu}|}{D_r} \cos\theta_i \right).
\label{equ:MFboltzmann}
\eeq
where the constant of proportionality is fixed by normalisation.
Combining (\ref{equ:MFnu},\ref{equ:MFboltzmann}) leads to a self-consistency relationship
\beq
|\bm{\nu}| = \int p^{\rm con}_{|\bm{\nu}| }(\theta )  \cos\theta \, {\rm d}\theta
\eeq
The integral can be expressed in terms of a Bessel function.  However, the relevant question is for which values of $g$ non-trivial solutions exist (excluding $|\bm{\nu}|=0$).  For that purpose one may expand for small values of the parameter $2 g |\bm{\nu}|/D_r$ which yields
$ \langle \cos\theta_i \rangle_{\rm con} = (g/D_r)|\bm{\nu}| + O(|\bm{\nu}|^3)$.  The correction term is negative and hence the non-trivial (ferromagnetic) solution appears for
\beq
g > D_r \; .
\label{equ:g>Dr}
\eeq

To analyse the paramagnetic phase we have by the central theorem of Sec.~\ref{sec:results-symm} that for $g=0$ then $p_0(\bm{\nu}) \propto \ee^{-N|\bm{\nu}|^2}$.   The Boltzmann distribution for the steady state of the controlled system is obtained by multiplying by $\ee^{-U_g^{\rm con}}$, yielding
\beq
p_g(\bm{\nu}) \propto \exp\left[-N |\bm{\nu}|^2 \left( 1 - g D_r^{-1} \right) \right]
\eeq
from which we obtain (\ref{equ:nu2-para}).
Consistent with the previous argument, that fluctuation diverges at the critical point $g=D_r$.
For larger $g$, estimation of $p_0$ by the central limit theorem is too simplistic and a more detailed analysis is required, for example as in (\ref{equ:MFnu},\ref{equ:MFboltzmann}).

\subsection{Large deviations of the time-averaged order parameter}
\label{app:pol}

In this section we consider large deviations of the time-averaged (vectorial) order parameter
\beq
\overline{\bnu}_\tau = \frac{1}{\tau} \int_0^\tau \bnu(t) dt \; .
\label{equ:average-nu}
\eeq
Recall that $\bnu(t)$ is defined in (\ref{equ:def-nu}) as a simple average of individual orientations which evolve independently by (\ref{equ:dr}).  Hence the statistics of $\overline{\bnu}_\tau$ can be analysed exactly.
As $\tau\to\infty$ there is a LDP
\beq
p( \overline{\bnu}_\tau ) \sim \exp[ -\tau N {\cal J}( |\overline{\bnu}_\tau| ) ]
\label{equ:LDP-nu}
\eeq
where ${\cal J}$ is the rate function, which only depends on the modulus of $\overline{\bnu}_\tau$, by symmetry. Note that the function $\cal J$ is distinct from ${\cal J}_1$ in (\ref{equ:LDP-nubar}), which describes large deviations of the time-integrated modulus of $\bnu$.  See however (\ref{equ:J1-J}), below.
There is an associated SCGF
\beq
\psi^{\rm OP}(h) = \lim_{\tau\to\infty} \frac{1}{N\tau} \log \left\langle \exp( - \tau N \bm{h} \cdot \overline{\bnu}_\tau  ) \right\rangle
\label{equ:psi-OP}
\eeq
where $h=|\bm{h}|$; the right hand side only depends on the modulus of $\bm{h}$, by symmetry.
There is a corresponding biased ensemble of trajectories, in which a generic observable $\cal A$ has average value
\beq
\langle {\cal A} \rangle_{\bm h} = \frac{ \langle {\cal A} \exp( -\tau N {\bm h} \cdot \overline{\bnu}_\tau ) \rangle }
{  \langle \exp( -\tau N {\bm h} \cdot \overline{\bnu}_\tau ) \rangle } \; .
\label{equ:bias-h}
\eeq

Since the rotors are independent under the ABP dynamics, the expectation value in (\ref{equ:psi-OP}) reduces to a product of expectation values for single rotors. Hence $\psi^{\rm OP}$ solves the eigenvalue problem
\beq
\psi^{\rm OP}(h) {\cal F}_h(\theta) = D_r {\cal F}_h''(\theta) - h  {\cal F}_h(\theta) \cos\theta
\label{equ:OPeigenproblem}
\eeq
As noted in~\cite{Grandpre2018}, this problem is related to Mathieu's equation.  Let $\tilde\theta = \theta/2$ and define $h$-dependent quantities $a=-4 \psi^{\rm OP}(h)/D_r$
and $q=2h/D_r$.
Defining also $\tilde{\cal F}(\tilde\theta) = {\cal F}_h(2\tilde\theta)$ we have
\beq
\tilde{\mathcal{F}}^{\prime\prime}(\tilde{\theta}) + (a - 2 q \cos 2 \tilde{\theta}) \tilde{\mathcal{F}}(\tilde{\theta}) = 0
\label{equ:Mathieu}
\eeq
We recognise Eq. \ref{equ:Mathieu} as Mathieu's differential equation \cite{Abramowitz1970}.  For any real number $q$ there is a countable infinity of possible values of $a$ and associated solutions $\tilde{\mathcal{F}}$. We are interested in  functions $\mathcal{F}_h$ that are even and $2\pi$-periodic.  Hence $\tilde{\cal F}$ must be $\pi$-periodic in $\tilde\theta$.   We therefore introduce $\mathcal{M}^{(0)}(\tilde\theta,q)$ which is the $0$\textsuperscript{th} even and $\pi$-periodic Mathieu function and $a^{(0)}_{\cal M}(q)$ its characteristic value \cite{Grandpre2018}, such that
\begin{align}
\label{equ:psiMathieu}
\psi^{\rm OP}(h) &= - \frac{D_r}{4} a^{(0)}_{\cal M}\left(2h/D_r\right),\\
\label{equ:fMathieu}
\mathcal{F}_h(\theta) &= \mathcal{M}^{(0)}\left(\frac{\theta}{2}, \frac{2h}{D_r} \right) .
\end{align}
Hence by Legendre transform the rate function in (\ref{equ:LDP-nu}) is
 \beq
 {\cal J}(\nubar) = \sup_h [ -h\nubar - \psi^{\rm OP}(h) ].
 \label{equ:JJ-legendre}
 \eeq
 This result is exact for all $N$ and all $\nubar$.

To obtain additional physical insight, we obtain the quadratic behaviour of the rate function at small $\nubar$.  This requires that we solve (\ref{equ:Mathieu}) for small $q$, which is a computation in perturbation theory~\cite{Arfken2013}.
Since $\tilde{\cal F}$ is even and $\pi$-periodic in $\tilde\theta$ it can be expanded as $\tilde{\cal F}(\tilde\theta) = 1 + \sum_{n=1}^\infty \beta_n \cos 2n\tilde\theta$.  The coefficients $\beta_n$ and the eigenvalue $a$ can then be expanded in powers of $q$.  To leading order,
\beq
a = -q^2/2 + O(q^4), \qquad \beta_1 = -q/2 + O(q^3)
\eeq
and $\beta_n=O(q^n)$ for $n\geq 2$.
Hence by (\ref{equ:psiMathieu},\ref{equ:fMathieu})
\beq
\psi^{\rm OP}(h) = \frac{1}{2 D_r} h^2 + O(h^4),
\label{equ:MathieuFExp}
\eeq
and so for small $\nubar$,
\beq
{\cal J}(\nubar) = \frac{1}{2} D_r \nubar^2 + O(\nubar^4) \; .
\label{equ:J-small-nu}
\eeq

The corresponding eigenfunction is
\beq
\mathcal{F}_h(\theta) = 1 - \frac{h}{D_r} \cos\theta + \mathcal{O}(h^2),
\eeq
Hence the optimal control potential (for this single orientation vector) is $U_{\rm opt}^{\rm OP}= - 2 \log {\cal F}_h$, so
\beq
U_{\rm opt}^{\rm OP}(\theta) 
= \frac{2h}{D_r}  \cos\theta + \mathcal{O}(h^2).
\label{equ:U-small-h}
\eeq
Recall, we are considering here large deviations where the order parameter is aligned parallel (or anti-parallel) the $x$-axis.  For $h>0$, the the control potential acts to align the orientation vectors anti-parallel to this axis, as expected from (\ref{equ:bias-h}).

\subsection{Time-averaged modulus of the order parameter}
\label{app:nu1}

\begin{figure}
\centering
\begin{overpic}[trim=0 0 0 0, width=0.48\textwidth]{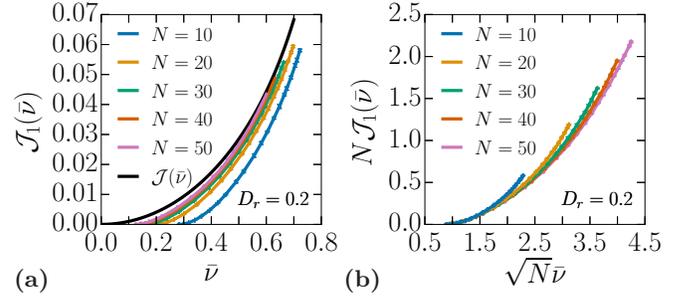}
\put (0, 0) {\texttt{\bf (a)}}
\put (\xp, 0) {\texttt{\bf (b)}}
\end{overpic}
\caption{{\bf (a, b)} Rate function of the time-averaged modulus of the order parameter $\mathcal{J}_1(\bar{\nu})$ computed from cloning simulations of rotors (\textit{Parameter values}: $n_c = 10^3$, $t_{\rm max} = 10^3$) {\bf (a)} and compared to the rate function of the time-averaged order parameter $\mathcal{J}(\bar{\nu})$ (See Eq.~\ref{equ:JJ-legendre}).}
\label{fig:rate-rotors}
\end{figure}

The discussion of the LDP (\ref{equ:LDP-nu}) of the previous section is useful as a way to characterise the LDP (\ref{equ:LDP-nubar}) of the main text.
In this case the relevant SCGF is
\beq
\psi_1(\lambda) = \lim_{\tau\to\infty} \frac{1}{N\tau} \log \left\langle \exp( - \tau N \lambda\nubar_\tau  ) \right\rangle
\label{equ:psi1}
\eeq
and ${\cal J}_1(\nubar) = \sup_\lambda [ -\lambda\nubar - \psi_1(\lambda) ]$.  In contrast to the previous case this problem cannot (to our knowledge) be solved exactly for finite $N$.  However as $N\to\infty$ the problem is of mean-field type.
In this case, the intuitive result is that the large-deviation mechanism for $\nubar_\tau$ should be the same as that of $\overline{\bnu}_\tau$, so that
\beq
\lim_{N\to\infty} {\cal J}_1(\nubar) = {\cal J}(\nubar)
\label{equ:J1-J}
\eeq
as illustrated by Fig.~\ref{fig:rate-rotors}(a).

We note from this figure that ${\cal J}$ has its zero at the origin, $\langle \overline{\nu}_{\tau} \rangle=0$.   On the other hand, (\ref{equ:para}) shows that the unique zero of ${\cal J}_1$ is at $\langle \overline{\nu}_{\tau} \rangle = \sqrt{\pi/(4N)}$ and in fact ${\cal J}_1$ increases rapidly for smaller values of $ \overline{\nu}_{\tau}$.  Hence, for any given $N$, there is a region between $0$ and $\langle \overline{\nu}_{\tau} \rangle$ where ${\cal J}_1$ deviates strongly from ${\cal J}$.  However, this region vanishes as $N\to\infty$ so (\ref{equ:J1-J}) holds for all $\nubar>0$.

Moreover, Appendix~F in Ref.~\cite{Nemoto2019} shows that
\beq
\tilde{\psi}_1(k) = \lim_{\tau \to \infty} \frac{1}{\tau} \log\left\langle e^{- k \int_0^{\tau} \sqrt{N} |\bm{\nu}(t)| \, \mathrm{d}t} \right\rangle
\eeq
is a well-defined smooth function for $N \gg 1$. Therefore, with $N \psi_1(\lambda) = \tilde{\psi}_1(\sqrt{N}\lambda)$, we obtain
\beq
N {\cal J}_1(\nubar) \simeq B \left(\sqrt{N}\nubar - \sqrt{\pi/4} \right)^2
\eeq
in the regime where $\sqrt{N} \nubar = \mathcal{O}(1)$, and where $B$ is a constant independent of $N$, as illustrated by Fig.~\ref{fig:rate-rotors}(b).

The next step is to
outline a derivation of \eqref{equ:J1-J}, which also yields the optimally-controlled dynamics for this problem.
The SCGF can be obtained by solving an eigenproblem
\beq
N\psi_1(\lambda) {\cal F}^{\rm pol}_{\lambda} = D_r\sum_i \frac{\partial^2}{\partial \theta_i^2} {\cal F}^{\rm pol}_{\lambda} - \lambda N|\bm{\nu}| {\cal F}_{\lambda}^{\rm pol}
\label{equ:eigen-1}
\eeq
where ${\cal F}_{\lambda}^{\rm pol} = {\cal F}_{\lambda}^{\rm pol}(\theta_1,\dots,\theta_N)$ depends on all angles.
Since this is a mean-field problem  the solution for large $N$ can be approximated as
\beq
{\cal F}_{\lambda}^{\rm pol} = f(\bnu) \prod_i \zeta_\lambda(\theta_i|\bnu)
\label{equ:F-lambda}
\eeq
where the orientation vectors interact only through their average, which is $\bnu$.  Note also $|\bm{\nu}| = N^{-1} \sum_i \cos(\theta_i-\varphi)$ where $\varphi$ is the angle between $\bm{\nu}$ and the $x$-axis.  We assume that $\zeta$ is normalised as $\int  \zeta(\theta|\bnu)^2 {\rm d}\theta= 1$.  The function $f$ is assumed to depend only on $|\bnu|$, and for large $N$ it is sharply-peaked in this variable, at $|\bm{\nu}|=\nu^*$.  In order that ${\cal F}^{\rm pol}$ is also sharply peaked at $\nu^*$, we require a self-consistency condition
\beq
\int  \cos(\theta-\varphi) \zeta(\theta|\bnu)^2 {\rm d}\theta = \nu^*
\label{equ:self-con1}
\eeq

The eigenproblem (\ref{equ:eigen-1}) is Hermitian (self-adjoint) so there is a variational (Rayleigh-Ritz) formula for its largest eigenvalue.  Using ${\cal F}^{\rm pol}$ as ansatz yields
\beq
 \psi_1(\lambda) \gtrsim -\lambda\nu^* - \frac{1}{2\pi N} \sum_i \int  \Psi_\zeta(\theta_i,\bnu^*)  {\rm d}\varphi
\label{equ:psi1-var}
\eeq
where we used (\ref{equ:self-con1}) as well as $\bm{\nu^*} = \nu^*(\cos\varphi,\sin\varphi)$ and
\beq
\Psi_\zeta(\theta,\bnu) = 
-\int D_r \zeta(\theta|\bm{\nu}) \frac{\partial^2}{\partial\theta^2} \zeta(\theta|\bm{\nu})  {\rm d}\theta_i \; .
\eeq
We have neglected terms in (\ref{equ:psi1-var}) which arise from the action of the derivatives on $\bm{\nu}$, these are negligible as $N\to\infty$.  In fact, the mean-field structure of the problem means that the bound (\ref{equ:psi1-var}) will become an equality as $N\to\infty$, if the optimal choice is made for $\zeta$.

It is convenient to work in terms of the rate function, using ${\cal J}_1(\nubar) = \sup_\lambda [ \lambda\nubar - \psi_1(\lambda)]$ to see that ${\cal J}_1(\nubar) \simeq \inf_\zeta \Psi_\zeta(\theta,\nubar)$ where the maximisation is again subject to (\ref{equ:self-con1}).  Implementing this constraint and the normalisation constraint on $\zeta$ by Lagrange multipliers $\mu_1,\mu_2$, we find
\beq
D_r \frac{\partial^2 \zeta}{\partial\theta^2} + [ \mu_1 - \mu_2\cos(\theta-\varphi) ] \zeta   = 0
\eeq
Hence we have recovered Mathieu's equation.
Proceeding similar to Appendix~\ref{app:pol} and using that extremisation of the Lagrange multipler $\mu_2$ yields a maximimum in this case, one obtains
\beq
{\cal J}_1(\nubar) \simeq \sup_{\mu_2} [  -\mu_2 \nubar - \psi^{\rm OP}(\mu_2) ].
\eeq
The notation $\simeq$ indicates that this relation becomes exact as $N\to\infty$.  Eq.~(\ref{equ:J1-J}) follows on comparing  with (\ref{equ:JJ-legendre}).
It follows that the optimal control potential for the LDP of (\ref{equ:LDP-nubar},\ref{equ:psi1}) is
\beq
U_{\rm con}^{\rm pol}(\theta_1,\dots,\theta_N) = -2\sum_i \log {\cal F}_{\lambda}\left({\theta_i-\varphi}\right)
\eeq
where the notation ${\cal F}_\lambda$ indicates the function defined in (\ref{equ:fMathieu}), evaluated at $h=\lambda$.  This is indeed a mean-field-type interaction among orientations.  By the same argument as (\ref{equ:U-small-h}), it reduces for small $\lambda$ to
\beq
U_{\rm con}^{\rm pol}(\theta_1,\dots,\theta_N) \simeq - \frac{2\lambda}{ND_r} \sum_{ij} \cos(\theta_i-\theta_j) \; .
\eeq
This is nothing but $U_g^{\rm con}$ from (\ref{equ:u-con-g}) with $g=2\lambda/D_r$.  The result is that for small values of $\nubar$, Eq.~(\ref{equ:u-con-g}) is an optimal control potential for large deviations of the orientation.

\subsection{Expansion of $\omega$}
\label{app:bias}

We assume that $\omega(\bar{\rho}, \bm{P})$ can be inferred from the ensemble of trajectories biased with respect to the polarisation $\bm{P}$ (Eq.~\ref{equ:biased-omega-hydro}),
\beq
\omega(\bar{\rho}, \bm{P}) = \left.\frac{\left\langle\bar{\rho} w_{\tau} \, e^{- \bm{h}(\bm{P}) \cdot \tau N \bar{\bm{\nu}}_{\tau}}\right\rangle}{\left\langle e^{- \bm{h}(\bm{P}) \cdot \tau N \bar{\bm{\nu}}_{\tau}}\right\rangle}\right|_{\bm{h}(\bm{P}),~ \left\langle\bar{\bm{\nu}}_{\tau}\right\rangle_{\bm{h}(\bm{P})} = \bm{P}}
\eeq
where we have used the averaged polarisation from Eq.~\ref{equ:average-nu}.

We have in the limit $\bm{h}(\bm{P}) \to \bm{0}$,
\begin{align}
&\begin{aligned} &\left\langle w_{\tau} \, e^{- \bm{h}(\bm{P}) \cdot \tau N \bar{\bm{\nu}}_{\tau}}\right\rangle\\ &\qquad = \left\langle w_{\tau} \right\rangle + \frac{1}{4} \tau^2 N^2 |\bm{h}(\bm{P})|^2 \left\langle w_{\tau} |\bar{\bm{\nu}}_{\tau}|^2\right\rangle,\end{aligned}\\
&\left\langle e^{- \bm{h}(\bm{P}) \cdot \tau N \bar{\bm{\nu}}_{\tau}}\right\rangle = 1 + \frac{1}{4} \tau^2 N^2 |\bm{h}(\bm{P})|^2 \left\langle |\bar{\bm{\nu}}_{\tau}|^2\right\rangle
\end{align}
up to $\mathcal{O}(\bm{h}(\bm{P})^2)$ terms and where we have discarded linear terms in $\bm{h}(\bm{P})$ by symmetry, therefore
\beq
\bar{\rho}^{-1} \omega(\bar{\rho}, \bm{P}) = \left\langle w_{\tau} \right\rangle + \frac{1}{4} \tau^2 N^2 |\bm{h}(\bm{P})|^2 \, \mathrm{Cov}(w_{\tau}, |\bar{\bm{\nu}}_{\tau}|^2)
\eeq
linking $\omega(\bar{\rho}, \bm{P})$ and the covariance of the active work and the squared average polarisation.

We note that
\beq
\left\langle\bar{\bm{\nu}}_{\tau}\right\rangle_{\bm{h}(\bm{P})} = - \frac{1}{2} \tau N \bm{h}(\bm{P}) \, \mathrm{Var}(\bar{\bm{\nu}}_{\tau})
\eeq
using $\left<\bar{\bm{\nu}}_{\tau}\right> = \bm{0}$ in the limit $\tau \to \infty$ such that $\mathrm{Var}(\bar{\bm{\nu}}_{\tau}) = \left\langle|\bar{\bm{\nu}}_{\tau}|^2\right\rangle$, and
\beq
\begin{aligned}
\left\langle|\bar{\bm{\nu}}_{\tau}|^2\right\rangle &= \frac{2}{D_r \tau N} \left(1 - \frac{1}{D_r \tau}\left(1 - e^{-D_r \tau}\right)\right)\\
&= \frac{2}{D_r \tau N},~ \tau \to \infty
\end{aligned}
\eeq
using $\langle \bm{u}(\theta_i(t)) \cdot \bm{u}(\theta_j(t^{\prime})) \rangle = \delta_{ij} e^{-D_r |t - t^{\prime}|}$ from Eq.~\ref{equ:dr}, therefore with $\left\langle\bar{\bm{\nu}}_{\tau}\right\rangle_{\bm{h}(\bm{P})} = \bm{P}$ we may write
\beq
\bm{h}(\bm{P}) = - \frac{2}{\tau N \mathrm{Var}(\bar{\bm{\nu}}_{\tau})} \bm{P} \underset{\tau\to\infty}{=} - D_r \bm{P},
\eeq
linking the biasing parameter and the polarisation.

We then have
\beq
\begin{aligned}
&\bar{\rho}^{-1} \omega(\bar{\rho}, \bm{P}) - \left\langle w_{\tau} \right\rangle = |\bm{P}|^2 \frac{1}{\mathrm{Var}(\bar{\bm{\nu}}_{\tau})^2} \mathrm{Cov}(w_{\tau}, |\bar{\bm{\nu}}_{\tau}|^2)\\
&\quad= \frac{1}{4} |\bm{P}|^2 \tau^2 \, N^2 \, D_r^2 \, \mathrm{Cov}(w_{\tau}, |\bar{\bm{\nu}}_{\tau}|^2),~ \tau \to \infty,
\end{aligned}
\eeq
at leading order in $|\bm{P}|^2$.

\section{Fluctuations of the active work in the hydrodynamic theory}
\label{app:fluctu-hydr}

This Appendix describes density fluctuations of ABPs at hydrodynamic level, including large deviations.  We follow Ref.~\cite{Dolezal2019}, which draws on earlier results including~\cite{Sollich2015,Appert2008}.
At hydrodynamic level, we are restricted to small biasing fields $s=O(1/L^2)$.  In this case, the (fast) polarisation field is unaffected by the bias and can be safely integrated out.    At the level of (\ref{equ:eom-field}), this leads to renormalisation of the diffusion constant $D_c$ but we do not distinguish the bare and renormalised values of $D_c$, for simplicity.   Since the polarisation has been integrated away, this analysis of density fluctuations is restricted to states with $\langle \bm{P}\rangle = 0$, but this is sufficient to cover homogeneous phases for $s>0$ (small enough that the system remains homogeneous) and for $s<0$ (small enough that there is no CM).

\subsection{Quadratic theory}

As in Refs.~\cite{Dolezal2019} (Section~5.1 and Appendix~B), we consider a perturbation around the homogeneous profile
\beq
\rho(\bm{r}, t) = \bar{\rho} + \delta\rho(\bm{r}, t)
\eeq
with $\delta\rho \ll \bar\rho$ and $\int \delta\rho(\bm{r},t) {\rm d}\bm{r}= 0$. Since we consider an isotropic system, $\bm{P} = \bm{0}$,  we define
\begin{align}
\bar{\omega}_0 &= \omega(\bar{\rho}, \bm{P} = \bm{0})\\
\bar{\omega}^{\prime\prime}_0 &= \frac{\partial^2}{\partial \rho^2} \omega(\bar{\rho}, \bm{P} = \bm{0})
\end{align}
so that we can Taylor expand the active work over $\rho$,
\beq
\begin{aligned}
N \tau w_{\tau} = &L^2 \tau \bar{\omega}_0 + \frac{1}{2} \bar{\omega}^{\prime\prime}_0 \int_{[0, L]^2} \int_0^{\tau} (\delta\rho)^2 \, \mathrm{d}^2\bm{r} \, \mathrm{d}t\\
&+ \mathcal{O}(\delta\rho^3).
\end{aligned}
\eeq
At the consistent level of expansion, the stochastic equation for the density (\ref{equ:eom-field}, \ref{equ:jd}) is
\beq
\frac{\partial}{\partial t} {\delta\rho} = D_c(\bar{\rho}) \nabla^2 \delta\rho - \sqrt{2 \sigma(\bar{\rho})} \nabla \cdot \bm{\eta}.
\label{equ:stoch-d-rho}
\eeq
We introduce the Fourier modes of the density,
\beq
\tilde{\rho}_{\bm{q}} = \frac{1}{L^2} \int_{[0, L]^2} \delta\rho(\bm{r}) e^{-\mathrm{i} \bm{q} \cdot \bm{r}} \, \mathrm{d}\bm{r}
\label{equ:rho-q}
\eeq
so that
\begin{align}
\delta \rho &=
\sum_{\bm{q} \neq (0, 0)} \tilde{\rho}_{\bm{q}} e^{\mathrm{i} \bm{q}\cdot\bm{r}} \; .
\end{align}
Hence
\beq
N  \tau  w_{\tau} = L^2 \tau \bar{\omega}_0 + L^2 \bar{\omega}^{\prime\prime}_0 \sum_{\substack{q_x \geq 0, q_y \\ \bm{q} \neq (0, 0)}} \int_0^{\tau} \tilde{\rho}_{\bm{q}} \tilde{\rho}_{-\bm{q}} \, \mathrm{d}t
\label{equ:w-in-modes}
\eeq
where the sum runs over non-zero modes $\bm{q}=2\pi(n_x,n_y)/L$, with $q_x \geq 0$, and where we have used
\beq
\int_{[0,L]^2} \delta\rho^2 \, \mathrm{d}^2\bm{r} = L^2 \sum_{\bm{q} \neq (0, 0)} \tilde{\rho}_{\bm{q}} \tilde{\rho}_{-\bm{q}} = 2 L^2 \sum_{\substack{q_x \geq 0, q_y \\ \bm{q} \neq (0, 0)}} \tilde{\rho}_{\bm{q}} \tilde{\rho}_{-\bm{q}}
\eeq
according to Parseval's theorem.   Since the theory is defined on the mesoscopic scale, sums over $\bm{q}$ are restricted to $|\bm{q}|<\Lambda$ where $\Lambda$ is an upper cutoff of order unity [to be precise, it is of order $|\Omega_{\bm{r}}|^{-1/d}$, for consistency with (\ref{equ:def-rho-field})].

From Eqs.~\ref{equ:stoch-d-rho} and \ref{equ:rho-q}, we derive the stochastic equation satisfied by the non-zero Fourier modes
\beq
\begin{aligned}
\frac{\partial}{\partial t} \tilde{\rho}_{\bm{q}} &= - D(\bar{\rho}) |\bm{q}|^2 \tilde{\rho}_{\bm{q}} + \sqrt{2\sigma(\bar{\rho}) |\bm{q}|^2} \tilde{\eta}_{\bm{q}}
\label{equ:eom-rhoq}
\end{aligned}
\eeq
where the longitudinal part of the noise term, $\tilde{\eta}_{\bm{q}}$, is a complex Gaussian white noise with zero mean and variance
\begin{align}
&\begin{aligned}
&\langle \tilde{\eta}_{\bm{q}}(t) \tilde{\eta}_{\bm{q}}^*(t^{\prime}) \rangle = \frac{1}{L^4 |\bm{q}|^2}\\
&\quad \times \iint_{[0, L]^2} \langle [(- \mathrm{i} \bm{q}) \cdot \bm{\eta}(t)] [(\mathrm{i} \bm{q}) \cdot \bm{\eta}(t^{\prime})] \rangle \, \mathrm{d}^2\bm{r} \, \mathrm{d}^2\bm{r}^{\prime}\\
&\quad = \delta(t - t^{\prime}),
\end{aligned}
\end{align}
and also
$
\langle \Re(\tilde{\eta}_{\bm{q}}(t)) \Im(\tilde{\eta}_{\bm{q}}(t^{\prime})) \rangle = 0.
$
Noises with different wavevectors $\bm{q}$ are independent.
The key point is that (\ref{equ:eom-rhoq}) is diagonal in $\bm{q}$ so every wavevector can be analysed separately.

\subsection{Biased ensemble of trajectories}

The next step is to consider a biased ensemble defined by the (linear) equations of motion (\ref{equ:eom-rhoq}) and the  reweighting factor $\ee^{-s\tau N w_\tau}$, where the exponential factor (\ref{equ:w-in-modes}) is quadratic (and diagonal) in the density fluctuations.

Following Ref.~\cite{Dolezal2019} (Appendix~B), we consider the complex Ornstein-Uhlenbeck process
\beq
z = - \zeta z + \sqrt{2 \gamma} \eta
\eeq
where $\eta$ is a complex Gaussian white noise with the same statistics as $\tilde{\eta}_{\bm{q}}$.
For a biased ensemble with exponential biasing factor of $\ee^{-sK_\tau}$ where $K_\tau=\alpha \int_0^\tau |z(t)|^2 dt$, the scaled cumulant generating function for $K_\tau$ can be computed.  Identifying $(z,\zeta,\gamma,\alpha)$ with $(\rho_{\bm{q}},D_c(\bar\rho)q^2,\sigma(\bar\rho)q^2,\omega_0'')$, the resulting SCGF for $w_\tau$ is  obtained by summing over the modes, to find
\begin{widetext}
\beq
\begin{aligned}
\psi(s) &= \lim_{\tau \rightarrow \infty} \frac{1}{N \tau} \log \left\langle e^{- s N \tau w_{\tau}}\right\rangle\\
&= - s \left\langle w_{\tau}\right\rangle - \frac{1}{N} \sum_{\substack{q_x \geq 0, q_y \\ \bm{q} \neq (0, 0)}} \left(\sqrt{D_c(\bar{\rho})^2|\bm{q}|^4 + 2 s \bar{\omega}_0^{\prime\prime} \sigma(\bar{\rho}) |\bm{q}|^2} - D_c(\bar{\rho})|\bm{q}|^2 - \frac{s \bar{\omega}_0^{\prime\prime} \sigma(\bar{\rho})}{D_c(\bar{\rho})}\right)
\end{aligned}
\label{equ:hydro-psi}
\eeq
\end{widetext}
consistent with \cite{Sollich2015, Appert2008}.
We emphasise that this result is valid only on the hydrodynamic scale, which means very small bias $s=O(1/L^2)$.

Several results are available from this formula. We first compute
\beq
-w'(s) = \psi''(s)
= \frac{[\omega_0^{\prime\prime} \sigma(\bar{\rho})]^2}{D(\bar{\rho})^3} \frac{1}{N} \sum_{\substack{q_x \geq 0, q_y \\ \bm{q} \neq (0, 0)}} \frac{1}{\bm{q}^2}
\eeq
which is
related to the variance of the active work from Eq.~\ref{equ:w'}.
The sum in this last expression can be approximated as
\beq
\begin{aligned}
\sum_{\substack{q_x \geq 0, q_y \\ \bm{q} \neq (0, 0)}} \frac{1}{\bm{q}^2} &
\simeq \frac{L^2}{(2\pi)^2}
\int_{2\pi/L}^{\Lambda}
 \frac{\pi q \, \mathrm{d}q}{q^2}\\
&= \frac{L^2}{4\pi} [ \log L + \mathcal{O}(1) ]
\end{aligned}
\eeq
where $\Lambda$ is the upper cutoff on $\bm{q}$.
Hence (\ref{equ:varW-logL}) follows, by combining these results with (\ref{equ:w'},\ref{equ:rhobar}).
The origin of this diverging variance is the presence of a slow (hydrodynamic) time scale, diverging as $L^2$.

The second result that is available from (\ref{equ:hydro-psi}) is that the argument of the square root will become negative if $s\omega_0''$ is sufficiently negative, indicating that the system becomes inhomogeneous. The instability is in the lowest mode, which has $|\bm{q}|=2\pi/L$.  Noting that $\omega_0''<0$ we obtain (\ref{equ:sc}), which is the point at which the system becomes unstable to phase separation.

Finally, observe that since $\psi$ in (\ref{equ:hydro-psi}) is the scaled cumulant generating function for squared density fluctuations, the structure factor of the biased ensemble can also be obtained by taking a derivative, leading to
\beq
\langle |\rho_{\bm q}|^2 \rangle_s = \frac{ \sigma(\bar\rho) |\bm{q}|^2 }{
  \sqrt{D_c(\bar{\rho})^2|\bm{q}|^4 + 2 s \bar{\omega}_0^{\prime\prime} \sigma(\bar{\rho})|\bm{q}|^2}
}
\label{equ:Sq-s}
\eeq
similar to~\cite{Sollich2015}.  Observe that the limiting behaviour of this function at $q\to0$ is different according to whether $s\omega_0''$ is zero or positive.  In the latter case then $\langle |\rho_{\bm q}|^2 \rangle_s\to0$ as $q\to0$, corresponding to hyperuniformity.  Hence (\ref{equ:Sq-HU}) follows.  If $s\omega_0''<0$ then it is not permissible to take $q\to0$ in (\ref{equ:Sq-s}), the argument of the square root would be negative at small $q$, which signals phase separation, as noted above.

\bibliography{references}

\end{document}